\pdfoutput=1

\documentclass[11pt,twoside,a4paper,cmspaper,final,collab]{cms-tdr}
\def\svnVersion{452059P}\def\cmsCernNoTag{CERN-EP-2017-201}\def\cmsCernDate{\today}\def\cmsMessage{Published in Physics Letters B as \href{http://dx.doi.org/10.1016/j.physletb.2018.03.028}{\doi{10.1016/j.physletb.2018.03.028.}}}

\begin{document}\cmsNoteHeader{SUS-16-042}

\hyphenation{had-ron-i-za-tion}
\hyphenation{cal-or-i-me-ter}
\hyphenation{de-vices}

\RCS$Revision: 446029 $
\RCS$HeadURL: svn+ssh://svn.cern.ch/reps/tdr2/papers/SUS-16-042/trunk/SUS-16-042.tex $
\RCS$Id: SUS-16-042.tex 446029 2018-02-14 15:21:18Z kirschen $
\newlength\cmsFigWidth
\newlength\cmsFigWidthB
\ifthenelse{\boolean{cms@external}}{\setlength\cmsFigWidth{0.90\columnwidth}}{\setlength\cmsFigWidth{0.48\textwidth}}
\ifthenelse{\boolean{cms@external}}{\setlength\cmsFigWidthB{0.80\textwidth}}{\setlength\cmsFigWidthB{0.90\textwidth}}

\ifthenelse{\boolean{cms@external}}{\providecommand{\cmsLeft}{upper\xspace}}{\providecommand{\cmsLeft}{left\xspace}}
\ifthenelse{\boolean{cms@external}}{\providecommand{\cmsRight}{lower\xspace}}{\providecommand{\cmsRight}{right\xspace}}
\ifthenelse{\boolean{cms@external}}{\providecommand{\CL}{C.L.\xspace}}{\providecommand{\CL}{CL\xspace}}
\ifthenelse{\boolean{cms@external}}{\providecommand{\NA}{\ensuremath{\cdots}\xspace}}{\providecommand{\NA}{\ensuremath{\text{---}}\xspace}}
\providecommand{\cmsTableResize[1]}{\resizebox{\textwidth}{!}{#1}}

\ifthenelse{\boolean{cms@external}}{\providecommand{\ptvecell}{\ensuremath{\vec{p}_\mathrm{T}^{\kern2pt\ell}}\xspace}}{\providecommand{\ptvecell}{\ensuremath{\vec{p}_\mathrm{T}^{\ell}}\xspace}}
\newcommand{\x}{\ensuremath{\phantom{0}}}
\newcommand{\y}{\ensuremath{\phantom{.}}}
\newcommand{\cmsPm}{\ensuremath{\,\pm\,}}
\newcolumntype{Y}{>{\centering\arraybackslash}p{2.5cm}}
\newcolumntype{Z}{>{\centering\arraybackslash}p{2.7cm}}

\newcommand{\LT}{\ensuremath{L_{\mathrm{T}}}\xspace}
\newcommand{\MT}{\ensuremath{M_{\mathrm{T}}}\xspace}
\newcommand{\MTt}{\ensuremath{M_{\mathrm{T2}}}\xspace}
\newcommand{\DF}{\ensuremath{\Delta \phi}\xspace}
\newcommand{\Rcs}{\ensuremath{R_{\mathrm{CS}}}\xspace}
\newcommand{\nbjet}{\ensuremath{n_{\PQb}}\xspace}
\newcommand{\nbtag}{\ensuremath{n_{\PQb}}\xspace}
\newcommand{\njet}{\ensuremath{n_{\text{jet}}}\xspace}
\newcommand{\njetSR}{\ensuremath{n_{\text{jet}}^{\mathrm{SR}}}\xspace}
\newcommand{\kappab}{\ensuremath{\kappa_{\PQb}}\xspace}
\newcommand{\kappaW}{\ensuremath{\kappa_{\PW}}\xspace}
\newcommand{\kappatt}{\ensuremath{\kappa_{\ttbar}}\xspace}
\newcommand{\Wjets}{\ensuremath{\PW\text{+jets}}\xspace}
\newcommand{\Zjets}{\ensuremath{\PZ\text{+jets}}\xspace}
\newcommand{\DYjets}{\ensuremath{\text{DY+jets}}\xspace}
\newcommand{\Rcscorr}{\ensuremath{R_{\mathrm{CS}}^{\text{data(corr)}}}}
\newcommand{\ttjets}{\ensuremath{\ttbar{\rm +jets}}\xspace}

\cmsNoteHeader{SUS-16-042} % This is over-written in the CMS environment: useful as preprint no. for export versions
\title{Search for supersymmetry in events with one lepton and multiple jets exploiting the angular correlation between the lepton and the missing transverse momentum in proton-proton collisions at \texorpdfstring{$\sqrt{s}=13\TeV$}{sqrt(s) = 13 TeV}}

\date{\today}

\abstract{
Results are presented from a search for supersymmetry in events with a single electron or muon and hadronic jets.
The data correspond to a sample of proton-proton collisions at $\sqrt{s}=13\TeV$ with an integrated luminosity of 35.9\fbinv,
recorded in 2016 by the CMS experiment.
A number of exclusive search regions are defined according to
the number of jets, the number of \PQb-tagged jets, the scalar sum of the transverse momenta of the jets,
and the scalar sum of the missing transverse momentum and the transverse momentum of the lepton.
Standard model background events are reduced significantly by
requiring a large azimuthal angle between the direction of the lepton and of
the reconstructed $\PW$ boson, computed under the hypothesis that all of the
missing transverse momentum in the event arises from a neutrino produced in the
leptonic decay of the $\PW$ boson.
The numbers of observed events are consistent with the expectations from
standard model processes, and the results are used to set lower limits on supersymmetric particle masses
in the context of two simplified models of gluino pair production.
In the first model, where each gluino decays to a top quark-antiquark pair and a neutralino, gluino masses up to 1.8\TeV are excluded at the 95\% CL. 
The second model considers a three-body decay to a light quark-antiquark pair and a chargino, which subsequently decays to a W boson and a neutralino. 
In this model, gluinos are excluded up to 1.9\TeV.
}

\hypersetup{
pdfauthor={CMS Collaboration},
pdftitle={Search for supersymmetry in events with one lepton and multiple jets exploiting the angular correlation between the lepton and the missing transverse momentum in proton-proton collisions at sqrt(s)=13 TeV},%
pdfsubject={CMS},
pdfkeywords={CMS, physics, supersymmetry}
}

\maketitle

\section{Introduction}\label{sec:Introduction}

Supersymmetry (SUSY)~\cite{Ramond:1971gb,Golfand:1971iw,Neveu:1971rx,
Volkov:1972jx,Wess:1973kz,Wess:1974tw,Fayet:1974pd,Nilles:1983ge} is a
promising extension of the standard model (SM) of particle physics. The
addition of supersymmetric partners to the SM particles can lead to the
suppression of quadratically divergent loop corrections to the mass squared of
the Higgs boson~\cite{Barbieri:1987fn}.  Furthermore, in SUSY models with
$R$-parity conservation~\cite{FARRAR1978575}, the lightest supersymmetric
particle (LSP) can provide a dark matter
candidate~\cite{Boehm:1999bj,Balazs:2004bu}.

This paper presents a search for SUSY in the single-lepton channel using data
recorded in 2016 by the CMS experiment at the CERN LHC, corresponding to an integrated
luminosity of 35.9\fbinv of proton-proton collisions at $\sqrt{s}=13\TeV$.
The analysis is an update of the search in Ref.~\cite{SUS-15-006},
which was performed using the significantly smaller data sample collected by CMS in 2015.
Similar searches were performed by the CMS and ATLAS experiments
at $\sqrt{s}=7\TeV$~\cite{Chatrchyan:2012ola,Chatrchyan:2012sca,Aad:2012ms},
8\TeV~\cite{Chatrchyan:2013iqa,Aad:2015mia,Aad:2014lra},
and 13\TeV~\cite{CMS-PAS-SUS-15-007,ATLAS-13TeV_single_lepton,ATLAS-13TeV_multib}.

\begin{figure}[!thb]
  \includegraphics[width=0.35\textwidth]{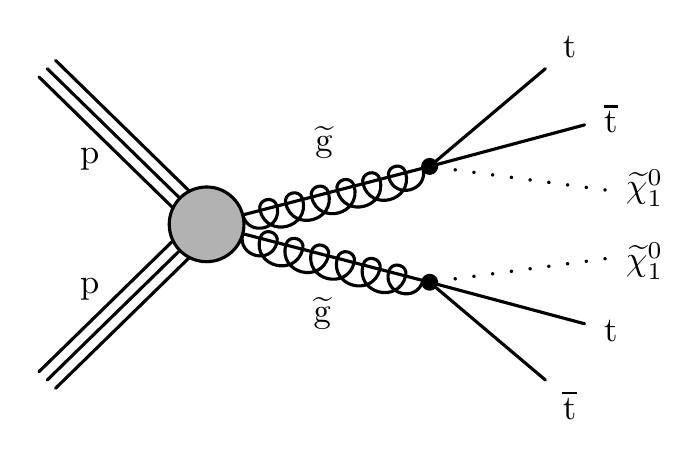}
  \includegraphics[width=0.35\textwidth]{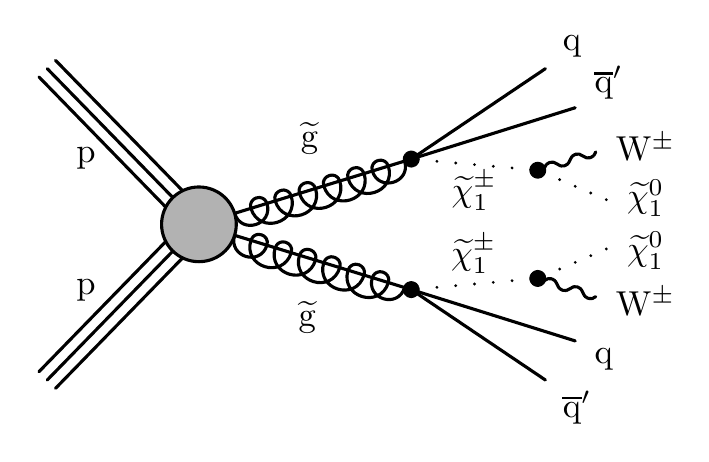}
\centering
  \caption{\label{fig:feynman_1} Diagrams showing the simplified models (\cmsLeft) T1tttt and
(\cmsRight) T5qqqqWW.
  }
\end{figure}

The results are interpreted within the framework of simplified
models~\cite{bib-sms-1,bib-sms-2,bib-sms-3,bib-sms-4} of gluino pair
production in which the LSP is the lightest neutralino, \PSGczDo, and
the lepton is produced in the decay of a W boson that originates either
from top-quark (\PQt) or chargino (\PSGcpmDo) decay.
In the T1tttt model shown in
Fig.~\ref{fig:feynman_1}~(\cmsLeft), gluinos (\PSg) undergo three-body
decays to $\ttbar + \PSGczDo$. In the T5qqqqWW model shown in
Fig.~\ref{fig:feynman_1}~(\cmsRight), the gluinos undergo three-body decays to a
first- or second-generation quark-antiquark pair (${\rm q}{\rm \bar{q}\prime}$) and a
\PSGcpmDo. The chargino is assumed to have mass ${m_{\PSGcpmDo}=0.5 
(m_{\PSg}+m_{\PSGczDo})}$ and to decay to a \PSGczDo and a $\PW$ boson.

\section{The CMS detector}\label{sec:detector}

The central feature of the CMS apparatus is a superconducting solenoid of
6\unit{m} internal diameter, providing a magnetic field of 3.8\unit{T}. Within
the solenoid volume are a silicon pixel and strip tracker, a lead tungstate
crystal electromagnetic calorimeter (ECAL), and a brass and scintillator hadron
calorimeter (HCAL), each composed of a barrel and two endcap sections. Forward
calorimeters extend the pseudorapidity ($\eta$) coverage provided by the barrel
and endcap detectors. Muons are detected in gas-ionization chambers embedded in
the steel flux-return yoke outside the solenoid.  A more detailed description
of the CMS detector, together with a definition of the coordinate system used
and the relevant kinematic variables, can be found in
Ref.~\cite{Chatrchyan:2008zzk}.  In what follows, the azimuthal angle around
the counterclockwise beam axis is denoted by $\phi$.

\section{Event reconstruction and simulation} \label{sec:reco}

The analysis makes use of the particle-flow event algorithm~\cite{Sirunyan:2017ulk},
which reconstructs and identifies each individual particle with an optimized combination of information
from the various elements
of the CMS detector. The energy of photons is directly obtained from the ECAL
measurement, corrected for zero-suppression effects. The energy of electrons is
determined from a combination of the electron momentum at the primary
interaction vertex as determined by the tracker, the energy of the
corresponding ECAL cluster, and the energy sum of all bremsstrahlung photons
spatially compatible with originating from the electron track. The energy of
muons is obtained from the curvature of the corresponding track. The energy of
charged hadrons is determined from a combination of their momentum measured in
the tracker and the matching ECAL and HCAL energy deposits, corrected for
zero-suppression effects and for the response function of the calorimeters to
hadronic showers. Finally, the energy of neutral hadrons is obtained from the
corresponding corrected ECAL and HCAL energy.

The degree of isolation of a lepton from other particles provides a strong indication of whether
it was produced within a jet, as would be expected from the fragmentation of a \PQb quark,
or in the leptonic decay of a $\PW$ boson, which can be produced either directly or in decays of
heavy particles such as the top quark.
The isolation is characterized by the scalar sum of the transverse momenta (\pt)
of all particles within a cone of radius $R = \sqrt{\smash[b]{(\Delta\eta)^2+(\Delta\phi)^2}}$
around the lepton momentum vector, excluding the contribution of the lepton and the contribution of charged particles not associated with the 
primary interaction vertex.
In the calculation of the isolation variable, an area-based correction is employed to remove
the contribution of particles from ``pileup''~\cite{Cacciari:2007fd},
\ie additional proton-proton collisions within the same or neighboring bunch crossings.
The isolation variable $I_\text{rel}$ is defined as the ratio of the scalar sum of the
\pt in the cone to the transverse momentum of the lepton, $\pt^{\ell}$.
To maintain high efficiency for signal events, which can contain a large number of
jets from the SUSY decay chains, a cone radius that depends on $\pt^{\ell}$, is used:
$R=0.2$ for $\pt^\ell < 50\GeV$, $10 / \pt^\ell[\GeVns{}]$ for $50 < \pt^\ell < 200\GeV$, and $0.05$ for $\pt^\ell > 200\GeV$.
This \pt dependent isolation definition additionally reduces the accidental overlap between jets and the lepton in regions where the SUSY decay products are boosted.
Accepted muons and electrons are required to satisfy
$I_\text{rel}<0.2$ and $I_\text{rel}<0.1$, respectively.

Jets are clustered using the anti-\kt algorithm~\cite{Cacciari:2008gp} with a distance parameter
of 0.4~\cite{Chatrchyan:2011ds}, as implemented in the \FASTJET package~\cite{Cacciari:2011ma}.
The momentum of a jet, which is determined as the vectorial sum of all particle momenta in the jet,
is found from simulation to be within 5 to 10\% of the true momentum over the full \pt spectrum
and detector acceptance. An offset correction is applied to jet energies
to take into account the contribution from pileup~\cite{Cacciari:2007fd}.
Jet energy corrections are derived from simulation and confirmed with in-situ
measurements of the energy balance in dijet, \Zjets, and photon+jet events~\cite{Khachatryan:2016kdb}.
Additional selection criteria are applied to each event to remove spurious jet-like features
originating from isolated noise patterns in certain HCAL regions. 
They have negligible impact on the efficiency for signal events. 
Jets originating from \PQb quarks are identified with an inclusive combined
secondary vertex tagging algorithm (CSVv2)~\cite{Chatrchyan:2012jua,Sirunyan:2017ezt}
that uses both secondary-vertex and track-based information.  The working point
is chosen to provide a \PQb tagging efficiency of ${\approx}63\%$, a \PQc tagging efficiency of 
${\approx}12\%$, and a light-flavor and gluon misidentification rate of ${\approx}0.9\%$ for 
jets with $\pt>20\GeV$ in simulated \ttbar events~\cite{Sirunyan:2017ezt}.
Double counting of objects is
avoided by not considering jets that lie within a cone of radius 0.4 around a
selected lepton.  To avoid double counting of objects as both a lepton and a
jet, jets that lie within a cone of radius $R = 0.4$ of a lepton are not
considered.

The missing transverse momentum vector, \ptvecmiss, is defined as the projection onto the plane perpendicular to the beam axis of the negative vector sum of the momenta of all reconstructed particle-flow objects in an event. 
Jet energy corrections are propagated to \ptvecmiss. Its magnitude is referred to as \ptmiss.

To estimate corrections to transfer factors extracted from data, and to
determine certain small backgrounds, Monte Carlo (MC) simulation is used. The
leading-order (LO) event generators {\MGvATNLO} v.2.2.2 or
v.2.3.3~\cite{Alwall:2014hca} are used to simulate \ttjets,
\Wjets, $\qqbar \to\Z/\gamma^* \to \ell^{+}\ell^{-}$ events, in the following referred to as \DYjets,
 and multijet events, in the following named QCD events. 
Events with a single top quark in the final state are generated using the
next-to-leading order (NLO) {\POWHEG}v2.0 and {\POWHEG}
programs~\cite{Nason:2004rx,Frixione:2007vw,Alioli:2010xd,Alioli:2009je,Re:2010bp}
for the $t$-channel and $\PQt\PW$ production, respectively.  The $s$-channel
single-top process and the production of both $\ttbar\PW$ and $\ttbar\PZ$,
commonly referred to as $\ttbar$V, are simulated using the NLO
{\MGvATNLO}  v.2.2.2
generator~\cite{Alwall:2014hca}.  The simulated background samples are
normalized using the most accurate cross section calculations
available~\cite{Alioli:2009je,Re:2010bp,Alwall:2014hca,Melia:2011tj,Beneke:2011mq,Cacciari:2011hy,Baernreuther:2012ws,Czakon:2012zr,Czakon:2012pz,Czakon:2013goa,Gavin:2012sy,Gavin:2010az},
which generally correspond to NLO or next-to-NLO (NNLO) precision.  All signal
events are generated with {\MGvATNLO} v.2.2.2, with up to two final-state partons in
addition to the gluino pair. {\MGvATNLO} uses the NNPDF3.0LO and the
NNPDF3.0NLO PDF~\cite{Ball:2014uwa} for processes with LO or NLO accuracy, respectively.  Gluino decays are based on
a unit matrix element~\cite{Sjostrand:2014zea}, with signal production cross
sections computed at NLO with next-to-leading-logarithm (NLL)
accuracy~\cite{bib-nlo-nll-01,bib-nlo-nll-02,bib-nlo-nll-03,bib-nlo-nll-04,bib-nlo-nll-05}.

Several benchmarks SUSY models, corresponding to different scenarios for the
gluino and neutralino masses, are used to study the kinematic properties of the
signal and to illustrate the numbers of events expected from SUSY.  The
benchmarks are denoted by the model name and the two key parameters, namely
$m_{\PSg}$ and $m_{\PSGczDo}$. As example, T1tttt(1.4, 1.1) corresponds to the
T1tttt model with $m_{\PSg}=1.4\TeV$ and $m_{\PSGczDo}=1.1\TeV$.  A second
benchmark, T1tttt(1.9, 0.1), is also used in this analysis.  Similarly, two
benchmark points are used to study the T5qqqqWW model: T5qqqqWW(1.9, 0.1) and
T5qqqqWW(1.5, 1.0). For the two T5qqqqWW benchmark models, the mass of the
intermediate chargino is taken to be 1.0\TeV and 1.25\TeV, respectively.

The evolution and hadronization of partons is performed using 
\PYTHIA~8.212 \cite{Sjostrand:2014zea} with the CUETP8M1
tune~\cite{Khachatryan:2015pea}.  Pileup is generated for a nominal
distribution in the number of pp interactions per bunch crossing, which is
subsequently reweighted to match the corresponding distribution observed in
data.  The detector response for all backgrounds is modeled using a detailed
simulation based \GEANTfour~\cite{Agostinelli:2002hh}, while a fast simulation
program~\cite{Abdullin:2011zz} is used to reduce computation time for signal
events.  The fast simulation has been validated against detailed
\GEANTfour-based simulations in reconstructed objects relevant to this search, 
and corresponding efficiency corrections based on data are applied to simulated 
background and signal events, respectively.

\section{Trigger and event selection}\label{sec:eventselection}

This analysis requires events containing a loosely isolated electron or muon 
with $\pt>15\GeV$ and a scalar sum of the jet transverse momenta in the event, 
$\HT$, with values greater than $400\GeV$ at the trigger level.
 To maximize the overall efficiency, additional
trigger paths were added requiring missing transverse momentum ($\ptmiss>100$,
110, or 120\GeV), isolated leptons ($\pt>27\GeV$
for electrons and $\pt>24\GeV$ for muons) or leptons with no isolation
requirement but with a higher \pt threshold ($\pt>105\GeV$ or $\pt>115\GeV$ for
electrons and $\pt>50\GeV$ for muons). The trigger efficiency is measured in
control samples recorded either with single-lepton triggers or with triggers with
 a requirement on \HT. After applying the
offline event selection requirements, an overall trigger efficiency of $(98\pm
1)\%$ is observed for the electron channel and negligible inefficiency for
the muon channel.

The event selection is similar to that presented in Ref.~\cite{SUS-15-006},
with improvements as noted to enhance the sensitivity of the analysis. Leptons
(electrons or muons) must satisfy $\pt > 25\GeV$.  Additional leptons with
$\pt>10\GeV$ that satisfy looser selection criteria of $I_\text{rel}<0.4$ are
referred to as ``veto'' leptons. To reduce the contribution from standard model
processes that produce higher lepton multiplicities, events with one or more
veto leptons are rejected.

Jets are required to have $\pt > 30\GeV$ and $|\eta| < 2.4$ to be considered for 
the calculation of higher level quantities such as \HT, the number of jets (\njet), 
and the number of \PQb-tagged jets (\nbtag). 
A number of exclusive kinematic regions, denoted as ``search bins'', are defined according to \njet, \nbtag, \HT, and the quantity $\pt^{\ell}+\ptmiss$ (\LT). All search bins are required
to contain at least five jets with the two highest-\pt jets satisfying $\pt > 80\GeV$.
Search bins with zero \PQb-tagged jets, called ``0-\PQb'', are mainly sensitive
to the T5qqqqWW model, while search bins with at least one \PQb-tagged jet,
called ``multi-\PQb'', are mainly sensitive to the T1tttt model. For the latter, the
requirement on the number of jets is increased to six, since the presence of four top quarks
results in an increased jet multiplicity in signal events.

To ensure that the analysis is sensitive both to signals with high \ptmiss as well as with
small \ptmiss but with large lepton \pt, no explicit threshold on \ptmiss is imposed.  Instead,
\LT is required to be ${>}250\GeV$.
Because of the trigger requirements and the extensive jet  activity expected in the chosen
SUSY models, \HT is required to be ${>}500\GeV$.

An important background arises from \ttjets events in which both $\PW$ bosons
decay leptonically and one lepton does not fulfill the selection criteria for
veto leptons.  In an extension of the previous analysis~\cite{SUS-15-006}, and
to suppress this background, events containing at least one isolated high-\pt
charged track are rejected in certain cases.  The high-\pt track can arise from $\tau\to
\nu_{\tau}+$hadron decays or muon or electron tracks of poor quality. The
relative isolation of such tracks within a cone of $R = 0.3$ around the track
candidate is required to be smaller than 0.1 or 0.2 for hadron or lepton
particle-flow candidates, respectively. For events containing such isolated
track candidates, the \MTt variable~\cite{Lester:mt2} is used: 
\begin{equation}
\ifthenelse{\boolean{cms@external}}
{
\begin{split}
  \MTt(\vec{p}_{\text T}^\ell,\vec{p}_{\text T}^t,\ptvecmiss) = & \\
  \min\limits_{\vec{p}_{\text T}^{(1)}+\vec{p}_{\text T}^{(2)} = \ptvecmiss} &
  \left\{ \max \left[ \MT(\vec{p}_{\text T}^\ell, \vec{p}_{\text T}^{(1)}),
\MT(\vec{p}_{\text T}^t,\vec{p}_{\text T}^{(2)})\right] \right\}, 
\end{split}
}
{
  \MTt(\vec{p}_{\text T}^\ell,\vec{p}_{\text T}^t,\ptvecmiss) =
  \min\limits_{\vec{p}_{\text T}^{(1)}+\vec{p}_{\text T}^{(2)} = \ptvecmiss}
  \left\{ \max \left[ \MT(\vec{p}_{\text T}^\ell, \vec{p}_{\text T}^{(1)}),
\MT(\vec{p}_{\text T}^t,\vec{p}_{\text T}^{(2)})\right] \right\}, 
}
\nonumber
\label{eq:MT2} 
\end{equation}  
where $\vec{p}_{\text T}^t$ and $\vec{p}_{\text
T}^\ell$ are the transverse momenta of the isolated track and the selected
lepton respectively,
and \MT is the transverse mass.  The minimization runs over all possible
splittings of \ptvecmiss assuming two lost massless particles, as in dileptonic
\ttbar decays that contain two neutrinos.  The isolated track with highest \pt
and opposite charge relative to the selected lepton is chosen where $\vec{p}_{\text T}^t$ is required to be ${>}5\GeV$. Events with a
hadronic or leptonic isolated track with \MTt below $60$ or $80\GeV$,
respectively, are rejected.  This requirement removes approximately 40\% of
dilepton \ttjets events, while rejecting only 8--15\% of the events in the SUSY
benchmark models.

\begin{figure*}[!htbp]
\centering
\includegraphics[width=\cmsFigWidth]{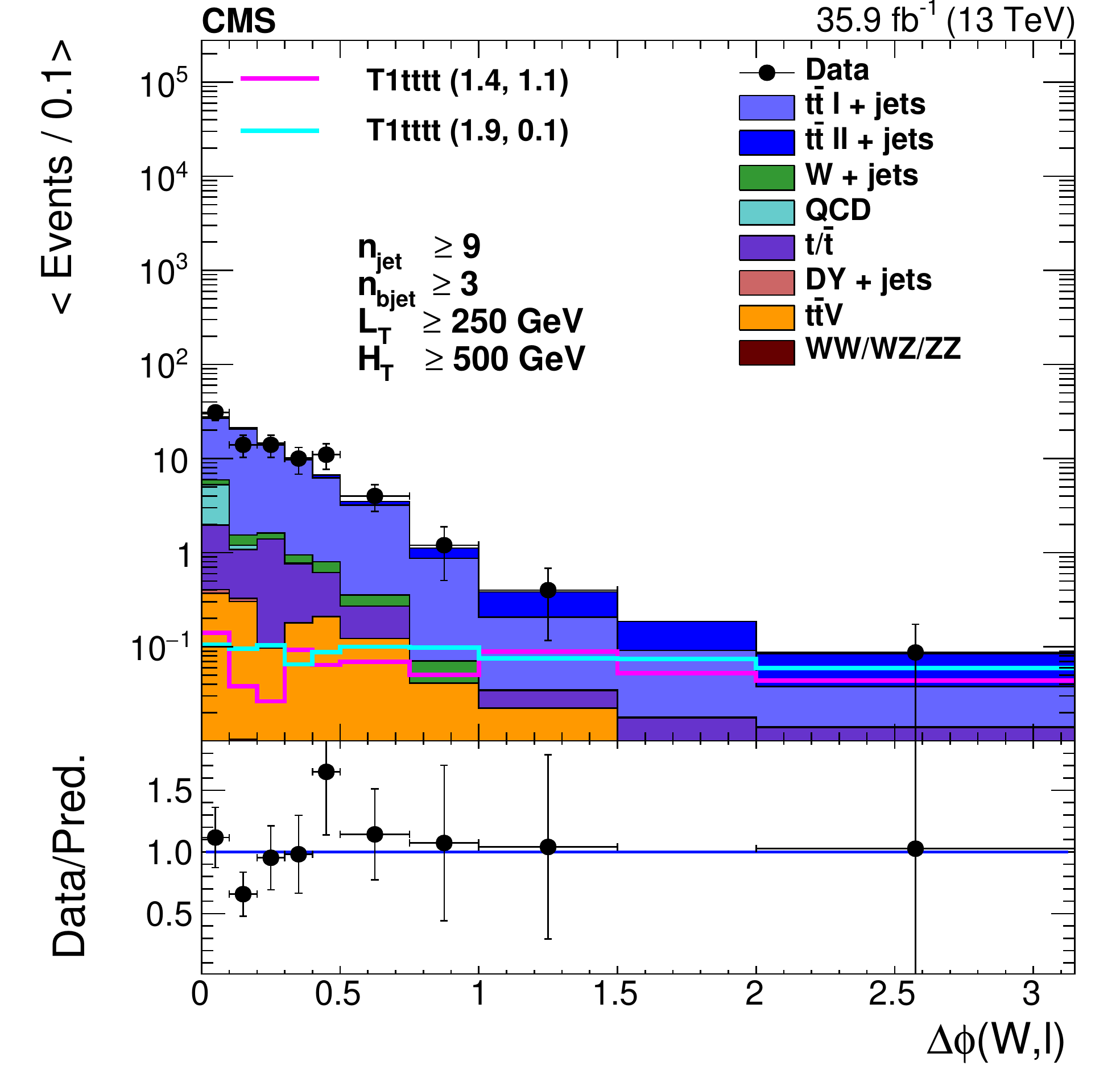} \hfil
\includegraphics[width=\cmsFigWidth]{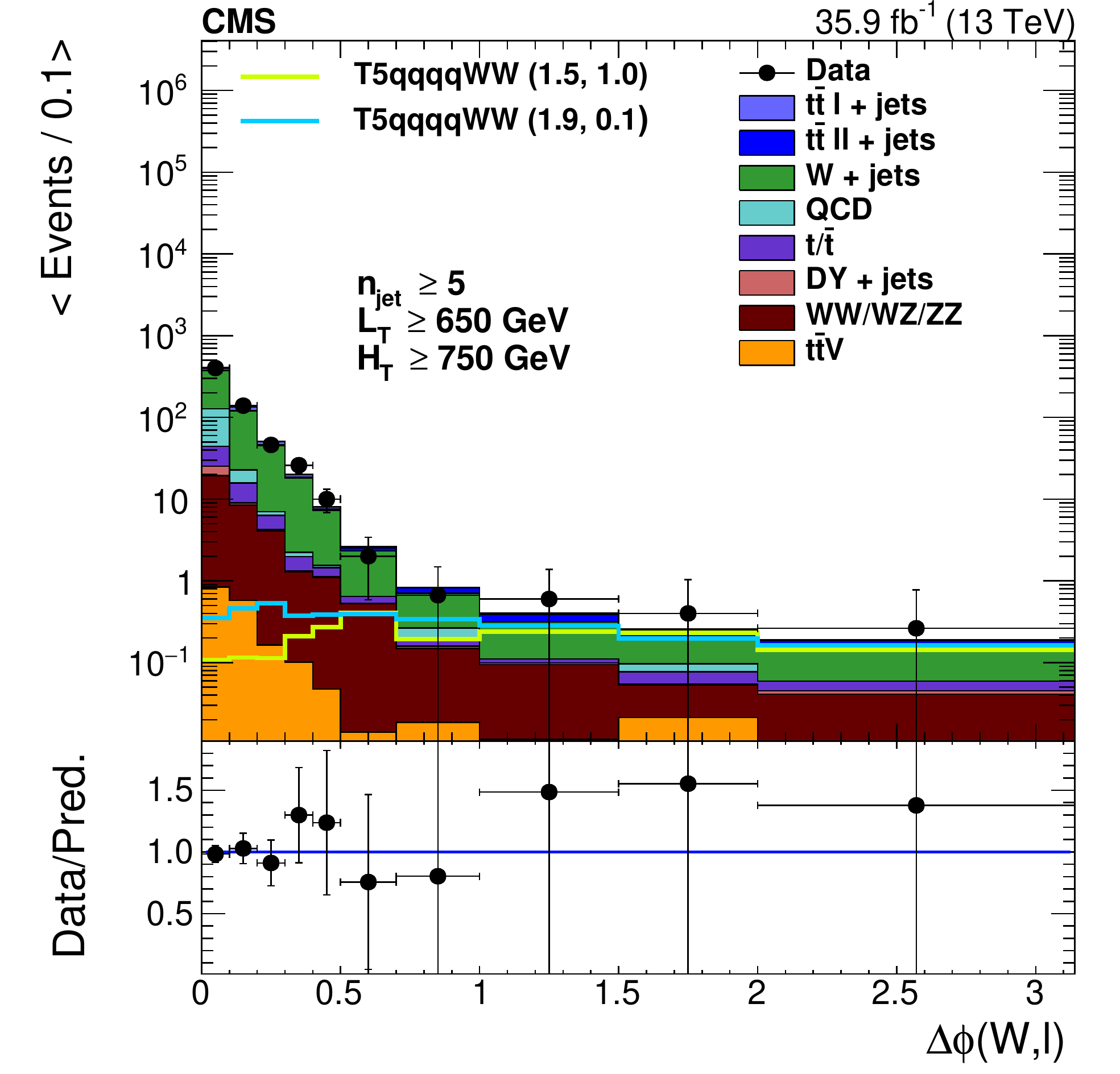}
\caption{Comparison of the \DF distribution for (left) the multi-\PQb and (right)
the 0-\PQb analysis for two of the search bins given in Table~\ref{tab:aggregated_regions}.
The simulated background events are stacked on top of each other and several signal
points are overlaid for illustration. The wider bins are normalized to
a bin width of 0.1. 
The ratio of data to simulation is given in the lower panels.
}
\label{fig:dPhi}
\end{figure*}

After these selections, the dominant remaining backgrounds are \Wjets events in
which the $\PW$ boson decays leptonically, and \ttjets events in which one of
the $\PW$ bosons from the top quarks decays leptonically and the other $\PW$
boson decays hadronically.  Both backgrounds are suppressed by requiring a
large azimuthal angle \DF between the lepton and the presumed $\PW$ boson. The
transverse momentum of the leptonically decaying $\PW$ boson is estimated as
the sum of \ptvecell and \ptvecmiss vectors.  In background events from \Wjets
and \ttjets with a single $\PW$-boson's leptonic decay, the \DF distribution
falls sharply and has a maximum value determined by the mass and \pt of the
$\PW$ boson. In the SUSY models investigated here, \ptvecmiss receives a large
contribution from the two neutralino LSPs. As a consequence, large values of
\DF are possible and the resulting \DF distribution in signal events is roughly
uniform.  The \DF variable can therefore be used to define the search region
(SR) as events with large \DF, while events with small \DF constitute the
control region (CR), which is used to estimate the SM background in the SR.
For illustration, Fig.~\ref{fig:dPhi} shows the \DF distributions in two
tightened multi-\PQb and 0-\PQb search bins as defined in
Table~\ref{tab:aggregated_regions}.  The magnitude of the angle between the
$\PW$ boson and the lepton is inversely proportional to the $\PW$ boson
momentum, which at high \pt is approximated by \LT. Therefore, the \DF
threshold used in defining the SR varies between 0.5 and 1, depending on
\LT.

The definitions of the search bins, along with the \DF values selected for the SRs,
are given in Tables~\ref{tab:PAS_table_multib} and \ref{tab:0b_results} for the multi-\PQb and
0-\PQb analyses, respectively. The name convention assigns a letter to each \njet and \nbtag category and a number from $0$ up to  $10$ for each \HT and \LT selection.
The multi-\PQb and the 0-\PQb analysis employ 39 and 28 search bins, respectively.

\section{Background estimation}\label{sec:background_estimation}

The method for estimating the background from SM processes is the same as the
one presented in Ref.~\cite{SUS-15-006}. For completeness, a summary of the
procedure is presented below.

The dominant backgrounds in all search bins arise from semileptonically
decaying \ttbar and leptonic \Wjets events. In each search bin, the number of
background events in the SR, \ie the yield of events at high \DF, is determined
using the number of events in the CR, \ie the events at low \DF, along with a
transfer factor $\Rcs$ that relates the events observed in the CR, $N_{\rm
data}(\rm CR)$, to those expected in the SR, $N_{\rm data}(\rm SR)$, as $\Rcs =
N_{\rm data}(\rm SR) / N_{\rm data}(\rm CR)$.

This transfer factor is measured in kinematic regions in data with a lower number of
jets, \njet, where the contribution from the signal is negligible.
Potential residual differences in transfer factors in the low- and high-\njet regions are determined
through simulation, where a correction factor, denoted by $\kappa$, is determined for each search bin as
$\kappa = \Rcs^{\rm MC}({\text{high-\njet}}) / \Rcs^{\rm MC}({\text{low-\njet}})$.

In the multi-\PQb analysis, the regions with one \PQb tag and four or five jets
consist of approximately 80\% \ttjets and 15--20\% \Wjets and single top quark events.
In all other multi-\PQb regions the \ttbar background is dominant. For this reason, only one transfer
factor is calculated  in the CRs with four or five jets to account for all backgrounds except QCD 
for each \LT, \HT and \nbtag range. This factor is then used to estimate the background in each 
SR of the search bins with $\njet \in [6-8]$ or $\njet \geq 9$.
A single transfer factor is used for the $\nbtag \ge 2$ search bins with the
same \HT and \LT, since these factors are found to be essentially independent of \nbtag.

In the 0-\PQb search bins, the contributions from \Wjets and \ttjets are
roughly equal, and a transfer factor for each background is determined in each
of the search bins in \njet, \HT, and \LT.
The transfer factor for \ttjets events is measured in data using events with
$\njet \in [4,5]$ and $\nbtag \geq 1$. For \Wjets events, the transfer factor
is measured also in data in events with $\njet \in [3,4]$ and $\nbjet = 0$; the
jet multiplicity used for \Wjets is lower than in \ttjets to limit the
contamination from \ttjets events.  The relative contribution of the \ttjets
and \Wjets components in the CR of each search bin is determined by a fit of
the \nbjet multiplicity distribution in the CR of the high-\njet regions, using
templates of the \nbjet multiplicity distributions for \Wjets and \ttjets that
are obtained from simulation. 
Additional backgrounds, including those from single top quark production, are found to 
be small and are taken from simulation.

About 10--15\% of the SM background events in the electron channel CRs are expected to be
QCD, and arise mainly from
jets misidentified as electrons or from photon conversions in the tracker. In
the SRs, however, the QCD background has been found to be negligible. It is estimated
from data, using ``antiselected'' events in which the electrons fail the
criteria for selected electrons but satisfy looser identification and isolation
requirements.  These events are scaled by the ratio of jets and photons that
pass the tight electron-identification requirements to the number of
antiselected electron candidates in a QCD-enriched sample that consists of no
\PQb-tagged jets and three or four jets.  To account for the QCD background in
the data, the QCD background is subtracted from the number of events in the CR
in the calculation of the transfer factor $\Rcs$ as well as from the number of events in the CR in each search bin. The prediction of the number of events in the SR of each search bin is then defined as:
\begin{equation}
N_{\rm pred}({\rm SR}) = \Rcs  \kappa  \left[ N_{\rm data}^{\rm high-\njet}({\rm CR}) - N^{\rm high-\njet}_{\rm QCD\; pred}({\rm CR}) \right]. \nonumber
\label{eq:prediction}
\end{equation}

The various $(\njet, \nbjet)$ regions employed in the analysis are described
in Table~\ref{tab:Regions}.

\begin{table*}[!htb]
\renewcommand{\arraystretch}{1.1}
\topcaption{Overview of the definitions of the various regions and samples employed in the analysis.
For the QCD fit the electron (e) sample is used, while for the determination (det.)
of $\Rcs(\Wpm)$ the muon ($\mu$) sample is used. Regions corresponding to blank cells are not used in the analysis.
}
\centering
\begin{tabular}{c|c|c|c|c}
Analysis       & \multicolumn{2}{c|}{Multi-\PQb analysis} & \multicolumn{2}{c}{0-\PQb analysis} \\ \hline
               & $\nbtag = 0$ & $\nbtag \geq 1$ & $\nbtag = 0$                & $\nbtag \geq 1$ \\ \hline
$\njet = 3$    &   QCD bkg.\ fit && $\Rcs(\Wpm)$ det. ($\mu$ sample), &        \\ \cline{1-1} \cline{3-3} \cline{5-5}
$\njet = 4$    &  (e sample) & \multirow{2}{*}{\Rcs det.} &  QCD bkg.\ fit (e sample) &  \multirow{2}{*}{$\Rcs(\ttjets)$ det.} \\  \cline{1-2} \cline{4-4}
$\njet = 5$    & & & \multirow{2}{*}{search bins} &  \\ \cline{5-5} \cline{1-3}
$\njet \geq 6$ & & search bins &   & \\
\end{tabular}
\label{tab:Regions}
\end{table*}

\section{Systematic uncertainties} \label{sec:systematics}

The systematic uncertainties are divided into two categories: those that affect the estimate of the
background from SM processes, and those that affect the expected signal yields.

The main systematic uncertainty on the background estimate arises from the uncertainty on
the value of the transfer factor \Rcs. The latter is measured in low-\njet data but is then
applied in the search bins that have higher jet multiplicities.
The modeling of jets from initial-state radiation (ISR) is obtained from a data sample
populated mainly by dilepton \ttjets events.
This sample is defined by two opposite-sign leptons (electrons or muons), excluding
events with same-flavor leptons within a window of $\pm 10\GeV$ around the $\PZ$-boson mass,
and two \PQb-tagged jets, such that any other remaining jets are interpreted as ISR.
In simulation, all jets that cannot be matched to daughter particles from the hard interaction are treated
as ISR jets.
The difference between the number of ISR jets observed and simulated is then used to
reweigh simulated \ttjets events in all analysis selections.
The reweighting factors vary between 0.92 and 0.51 for $N_\mathrm{J}^\mathrm{ISR}$ between 1 and 6.
We take one half of the deviation from unity as the systematic uncertainty on
these reweighting factors.

The presence of two neutrinos in dilepton \ttjets events tends to produce
larger angles between the lepton and the presumed $\PW$ boson than in
single-lepton \ttjets events.  As a result, the fraction of dilepton \ttjets in
which the second lepton does not pass the veto lepton requirements, is larger
at high \DF values, \ie in the SR, than in the CR.  This fraction as a function
of \njet must be described well in the simulation, as the differences in the
transfer factors between the low-\njet and high-\njet events, \ie the $\kappa$
factors, are determined in simulation.  This assumption is tested using
dilepton events, selected as described in the previous paragraph and split into
a 0-\PQb and a multi-\PQb category.  To study the behavior of the background
from dilepton events that remain in the single-lepton selection because of the
loss of one lepton, one of the two leptons is removed from the event. Since in
this type of background, the lost leptons arise principally from $\tau
\to\text{hadrons} + \nu$ decays, and to account for the \ptmiss due to the
neutrino from the $\tau$ decay, the lepton removed is replaced by a jet with
2/3 of the \pt of the original lepton and the \LT, \DF, and \HT values are
recalculated for the resulting ``single-lepton'' event.  To maximize the number
of events in the dilepton \ttjets control sample, no \DF requirement is
applied, and all events are used twice, with each reconstructed lepton
considered as the lost lepton.  The jet multiplicity in the single-lepton
baseline selection (excluding the SR) is compared with that in the
corresponding simulated event sample. In addition, the jet multiplicity in the dilepton \ttjets control sample in data is
compared with the corresponding simulated event sample. From these two comparisons a
double-ratio is formed. The remaining differences in the double-ratio, which are of the
order of 3--6\% per \njet bin, are corrected through the calculated $\kappa$
factors, and propagated as a systematic uncertainty.

Uncertainties in the background estimate that also affect the signal
arise from uncertainties in the jet energy scale (JES)~\cite{Chatrchyan:2011ds}, from uncertainties 
in the scale factors correcting the efficiencies and misidentification rate for 
\PQb tagging~\cite{Sirunyan:2017ezt}, and from uncertainties in the reconstruction and 
identification efficiencies of leptons~\cite{Khachatryan:2015hwa,Chatrchyan:2012xi}.

 In each case,
the systematic uncertainty in the background is estimated by changing the corresponding
correction factors within their uncertainties. After each such change in the JES,
the \HT and \ptmiss in each event are recalculated.
Similarly, the uncertainty arising from pileup is estimated by varying the inelastic cross section 
by its 5\% uncertainty~\cite{Aaboud:2016mmw}.

The \Wjets and \ttjets cross sections are varied independently by
30\%~\cite{Khachatryan:2016ipq} to account for possible biases in the
estimation of the background composition in terms of \Wjets \vs \ttjets events,
which changes slightly the value of $\kappa$.  These changes have only a small
impact on the 0-\PQb analysis, where the relative fraction of the two processes
is determined from a fit. In the multi-\PQb analysis, the differences in the
$\kappa$ values of less than 3\% are propagated to the background estimates.
The $\ttbar$V cross section is varied by 100\%. The systematic uncertainty in
the QCD background depends on \njet and \nbtag, and ranges from 25\% up to
100\% for the highest \nbtag region.

The polarization of $\PW$ bosons is changed by reweighting events by the factor
$w(\cos{\theta^*})= 1 + \alpha(1-\cos{\theta^*})^2$, where $\theta^*$ is the angle between
the charged lepton and $\PW$ boson in the $\PW$ boson rest frame.
For \Wjets events, we use $\alpha=0.1$, guided by the measurements and
theoretical uncertainties~\cite{Bern:2011ie,Khachatryan:2015paa,Chatrchyan:2011ig,ATLAS:2012au}.
For \ttjets events, we use $\alpha= 0.05$~\cite{Aad:2012ky,Khachatryan:2015dzz,Aaboud:2016hsq,Czarnecki:2010gb}.
For \Wjets events, where the initial state can have different polarizations for
$\PWp$ and $\PWm$ bosons, the  uncertainty is determined by the larger change in $\kappa$
resulting from reweighting only the $\PWp$ bosons in the sample, and from reweighting
all $\PW$ bosons. 

For the 0-\PQb analysis, an additional systematic uncertainty is based on 
linear fits of \Rcs as a function of \njet that are found to describe the dependence 
within statistical uncertainties.
A 50\% cross section uncertainty is used for all backgrounds other than \Wjets, \ttjets, $\ttbar$V, and QCD.

For the signal, an uncertainty in ISR is applied using
the approach described previously for the reweighting of the distribution of ISR jets in \ttjets as both, signal and \ttjets, 
rely on {\MGvATNLO} for event generation.
Half of the correction is used as an estimate of the uncertainty as is propagated to the signal acceptance.
To gauge their impact, the factorization and renormalization scales are changed up and down by a factor of 2.

Finally, the luminosity is measured using the pixel cluster counting method~\cite{CMS-PAS-LUM-15-001}, with the
absolute luminosity obtained using Van der Meer scans.
The resulting uncertainty is estimated to be 2.5\%~\cite{CMS-PAS-LUM-17-001}.

The impact of the systematic uncertainties on the estimate of the total background in
the multi-\PQb and 0-\PQb analyses is summarized in Table~\ref{tab:sysTable}.
While systematic uncertainties are determined for each signal point, typical values
for most signals are summarized for illustration in Table~\ref{tab:sysSigTab}.

\begin{table}[!htb]
\centering
\topcaption{Summary of systematic uncertainties in the total background estimates for
the multi-\PQb and for the 0-\PQb analyses.}
\label{tab:sysTable}
\ifthenelse{\boolean{cms@external}}{
\resizebox{\columnwidth}{!}{
\begin{tabular}{l@{}c@{}c}
Source                  & 
\begin{tabular}{c} Uncertainty \\ for multi-$\PQb$ \\ {[\%]} \end{tabular} & 
\begin{tabular}{c} Uncertainty \\ for 0-\PQb \\ {[\%]}   \end{tabular}  
\\ \hline
Dilepton control sample & 0.9--7.0                        & 0.3--18\y     \\
JES                     & 0.3--18\y                         & 0.7--26\y     \\
Tagging of \PQb jets    & 0.1--0.9     &     0.1--2.5         \\
Mistagging   & \multirow{2}{*}{0.1--2.2}    &    \multirow{2}{*}{0.3--0.8}         \\
\quad of light flavor jets   &     &            \\
$\sigma(\Wjets)$        & 0.3--9.3                        & 0.3--10\y     \\
$\sigma(\ttbar)$        & 0.1--7.5                        & 0.7--13\y     \\
$\sigma(\ttbar$V)       & 0.2--20\y                         & 0.1--3.8    \\
$\PW$ polarization          & 0.1--3.3                        & 0.7--14\y   \\
ISR reweighting (\ttbar)        & 0.5--7.0                        & 0.2--11\y     \\
Pileup                  & 0.4--7.1                        & 0.1--20\y   \\
Statistical uncertainty &  \multirow{2}{*}{\x5--30}                           & \multirow{2}{*}{\x5--36}   \\  
\quad  in MC events  & & \\
\hline
\end{tabular}
}}
{
\begin{tabular}{lcc}
Source                  & 
Uncertainty for multi-$\PQb$ [\%]  & 
Uncertainty for 0-\PQb [\%]    
\\ \hline
Dilepton control sample & 0.9--7.0                        & 0.3--18\y     \\
JES                     & 0.3--18\y                         & 0.7--26\y     \\
Tagging of \PQb jets    & 0.1--0.9     &     0.1--2.5         \\
Mistagging  of light flavor jets & 0.1--2.2    &    0.3--0.8         \\
$\sigma(\Wjets)$        & 0.3--9.3                        & 0.3--10\y     \\
$\sigma(\ttbar)$        & 0.1--7.5                        & 0.7--13\y     \\
$\sigma(\ttbar$V)       & 0.2--20\y                         & 0.1--3.8    \\
$\PW$ polarization          & 0.1--3.3                        & 0.7--14\y   \\
ISR reweighting (\ttbar)        & 0.5--7.0                        & 0.2--11\y     \\
Pileup                  & 0.4--7.1                        & 0.1--20\y   \\
Statistical uncertainty in MC events &  \x5--30                           &  \x5--36   \\  
\hline
\end{tabular}
}
\end{table}

\begin{table}[!h]
\centering
\topcaption{Summary of the systematic uncertainties and their average effect on the
yields for the benchmark points defined in the text. The values, which are quite similar for
the multi-\PQb and the 0-\PQb analyses, are usually larger for compressed scenarios,
where the mass difference between the gluino and the lightest neutralino is small.}
\label{tab:sysSigTab}
\ifthenelse{\boolean{cms@external}}{\resizebox{\columnwidth}{!}}{}
{
\begin{tabular}{lc}

Source                                  & Uncertainty [\%]      \\ \hline
Trigger                                 & 2                     \\
Pileup                                  & 10                    \\
Lepton efficiency                       & 2                     \\
Isolated track veto                     & 4                     \\
Luminosity                              & 2.5                   \\
ISR                                     & \x2--25                 \\
Tagging of \PQb jets                    & 1--6                  \\
Mistagging of light flavor jets         & 1--4                  \\
JES                                     & \x3--40                 \\
Factorization/renormalization scale     & 1--3                  \\
\ptmiss                                    & \x2--20                 \\ \hline
\end{tabular}
}
\end{table}

\section{Results and interpretation}\label{sec:results_interpretation}

\begin{figure*}[tbp!]
\centering
\includegraphics[width=\cmsFigWidthB]{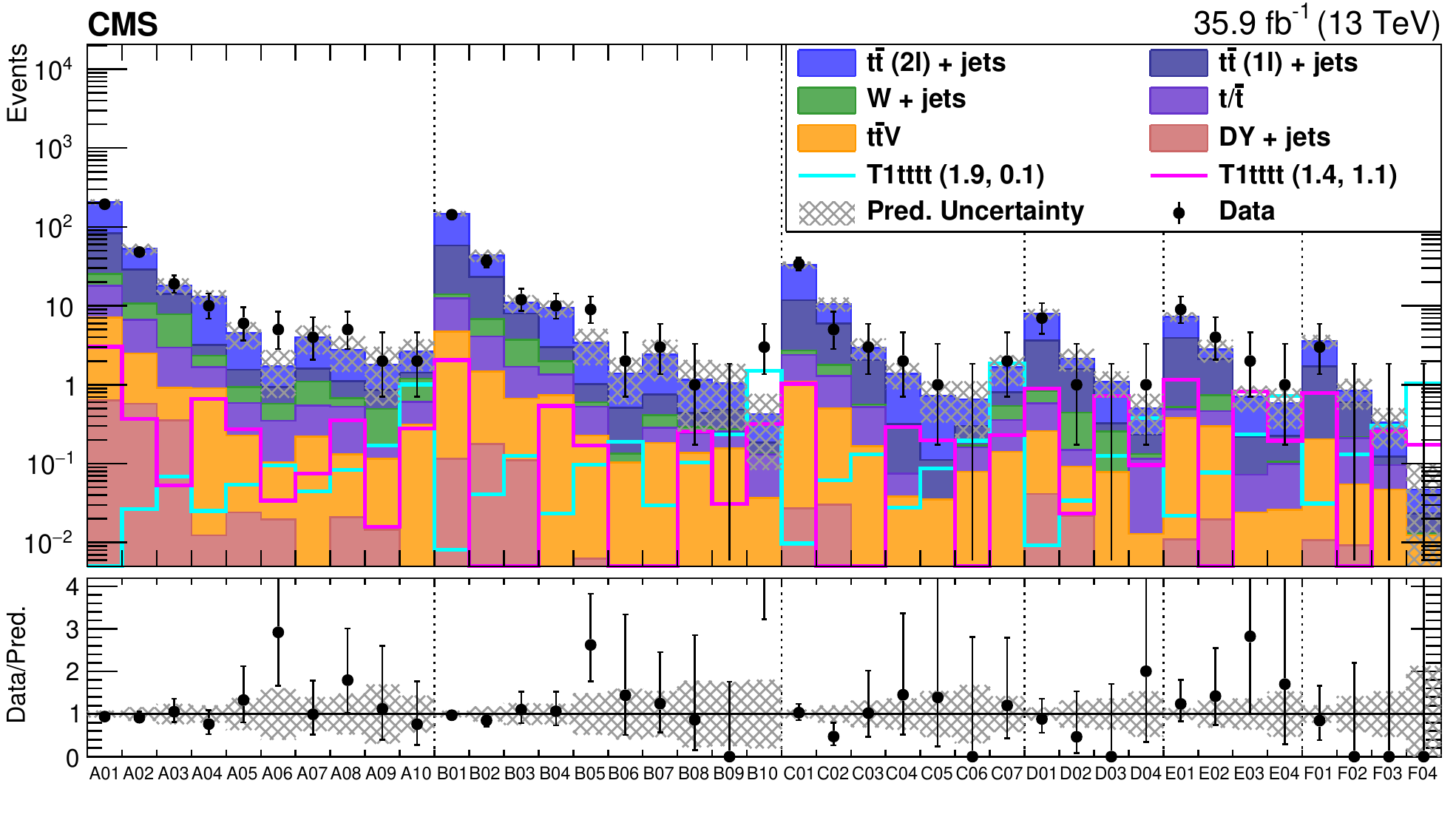}
\caption{
Multi-\PQb search: comparison of the numbers of events observed in the data and
the numbers expected from the estimated SM backgrounds in the 39 search bins
defined in the text, with details given in Table~\ref{tab:PAS_table_multib}.
Upper panel: the data are represented by black points with error bars, while the total
SM background expected is shown as a hatched region that represents
the uncertainty. For illustration, the relative fraction of the different SM
background contributions determined in simulation is shown by the stacked,
colored histograms, normalized so that their sum is equal to the
background estimated using data control regions, as described in the text.
The expected event yields for two T1tttt SUSY benchmark models are represented
by the open histograms.
Lower panel: the ratio of the number of events observed in data to
the number of events expected from the SM background in each search bin.
The error bars on the data points indicate the statistical uncertainty in the ratio, while the gray hatched region indicates the uncertainty on
this ratio from the uncertainty in the background estimate.
}
\label{fig:multibunblind}
\end{figure*}

The data in the search regions are compared to the background estimates in
Figure~\ref{fig:multibunblind} for the multi-\PQb events, where the outline of the
filled histogram represents the total estimated number of
background events. For illustration, the expected composition of the background is 
shown, assuming the relative fractions of the different SM processes (\ttjets, \Wjets, 
and other backgrounds), as determinated from simulation.

\begin{figure*}[tbp!]
\centering
\includegraphics[width=\cmsFigWidthB]{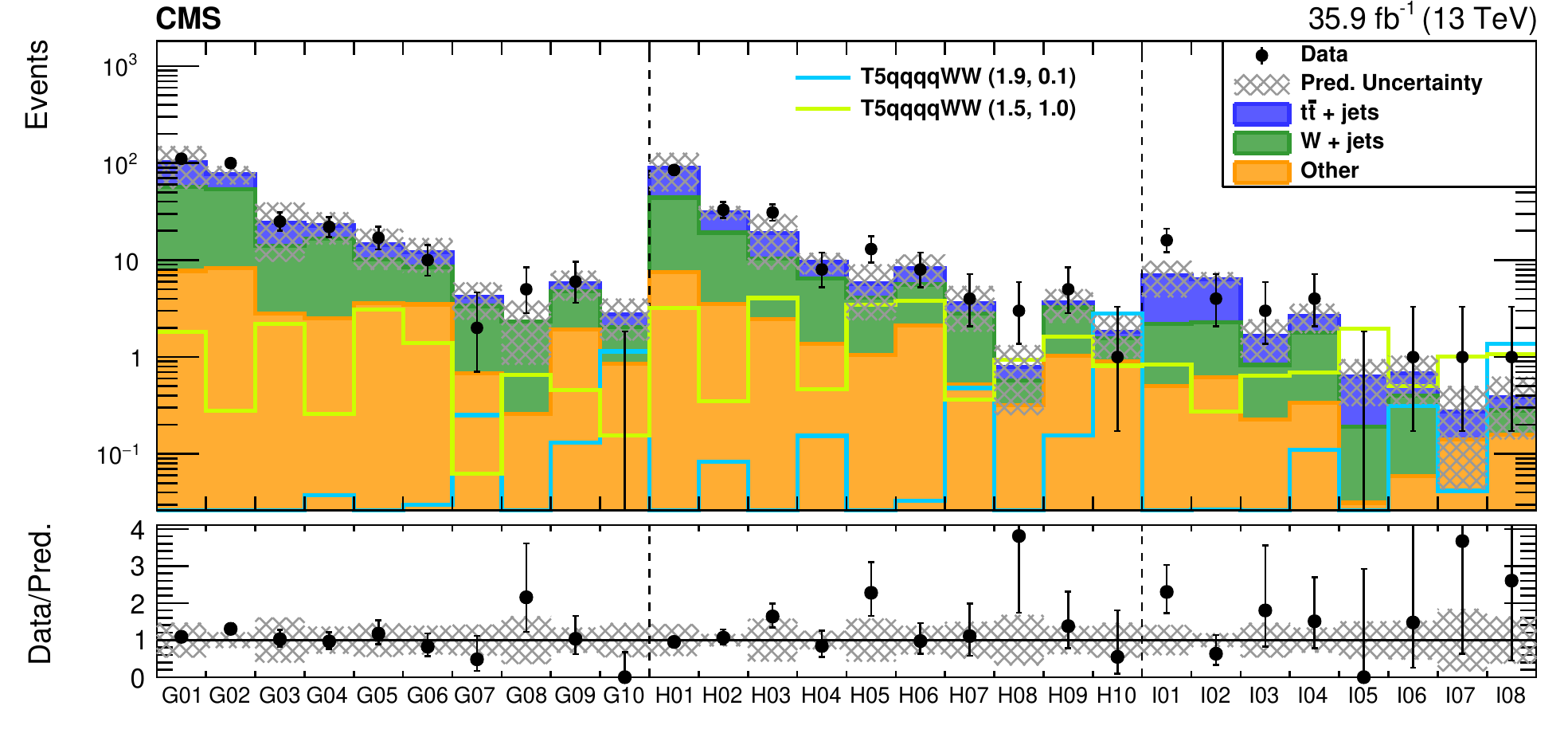}
\caption{0-\PQb search: comparison of the numbers of events observed in the data and
the numbers expected from the estimated SM backgrounds in the 28 search bins
defined in the text, with details given in Table~\ref{tab:0b_results}.
Upper panel: the data are represented by black points with error bars, while the total
SM background expected is shown as a hatched region that represents
the uncertainty.
The filled, stacked histograms represent the predictions for \ttjets, \Wjets events,
and the remaining backgrounds.
The expected yields from two T5qqqqWW SUSY benchmark models are represented as solid lines.
Lower panel: the ratio of the number of events observed in data to
the number of events expected from the SM background in each search bin.
The error bars on the data points indicate the statistical
uncertainty in the ratio, while the gray hatched region indicates the uncertainty on
this ratio from the uncertainty in the background estimate.
}
\label{fig:zerobunblind}
\end{figure*}

Figure~\ref{fig:zerobunblind} displays the estimates and data observed in the
0-\PQb events. The filled histogram represents the estimates
from data for \ttjets and \Wjets events and the remaining backgrounds,
which include the QCD estimate determined from data and rare backgrounds
determined from simulation.

To facilitate the reinterpretation of the results in terms of models not
considered here, a comparison of the background estimates and the observed
number of events in the SR of a few aggregated search bins is
presented in Table~\ref{tab:aggregated_regions}.  The results for all bins,
compared to two benchmark points, are given in
Tables~\ref{tab:PAS_table_multib} and \ref{tab:0b_results} for the multi-\PQb
and 0-\PQb analyses, respectively.  The data agree with the expectations from
the SM and no significant excess is observed.

\begin{table*}[!ht]
\topcaption{Definition of search bins and naming convention in the multi-\PQb
  search. Also given are the \DF values that are used to define the CRs and
  the SRs, the numbers of expected background events with combined statistical
  and systematic uncertainties, the observed numbers of events, and the
expected numbers of signal events in the multi-\PQb search bins.}
\centering
\label{tab:PAS_table_multib}
\cmsTableResize{
\begin{tabular}{c | c | c | c | c | l | Y | Y | Y | c }

\multirow{2}{*}{\njet} & \multirow{2}{*}{\nbjet} &\LT   & \DF    &   \HT  & \multicolumn{1}{c|}{Bin} &
    \multicolumn{2}{c|}{Signal T1tttt ($m_{\PSg}$, $m_{\PSGcz}$) $[$TeV$]$} & Predicted & Observed  \\
   &   & [GeV]  & [rad] &  [GeV]  &   \multicolumn{1}{c|}{name}      & \multicolumn{1}{c|}{(1.9, 0.1)}      &            \multicolumn{1}{c|}{(1.4, 1.1)}         &            background                &  data \\ \hline

\multirow{26}{*}{[6, 8]} & \multirow{9}{*}{${=}1$} & \multirow{3}{*}{[250, 450]} & \multirow{3}{*}{1.0} &  \x[500, 1000]& \y A01 & ${<}0.01$ & 3.02 $\cmsPm$ 0.24 & 206 $\cmsPm$ 15\x & 194 \\
&  & & & [1000, 1500] &\y A02 & 0.03 $\cmsPm$ 0.01 & 0.37 $\cmsPm$ 0.08 & 52.5 $\cmsPm$ 8.2\x & 48 \\
&  & & & $\geq$1500 &\y A03 & 0.07 $\cmsPm$ 0.01 & 0.05 $\cmsPm$ 0.03 & 18.0 $\cmsPm$ 4.2\x & 19 \\
\cline{3-10} &   & \multirow{3}{*}{[450, 600]} & \multirow{3}{*}{0.75} & \x[500, 1000]&\y A04 & 0.03 $\cmsPm$ 0.01 & 0.66 $\cmsPm$ 0.11 & 13.1 $\cmsPm$ 2.7\x & 10 \\
&  & & & [1000, 1500] &\y A05 & 0.05 $\cmsPm$ 0.01 & 0.27 $\cmsPm$ 0.07 & 4.5 $\cmsPm$ 1.7 & 6 \\
&  & & & $\geq$1500 &\y A06 & 0.09 $\cmsPm$ 0.01 & 0.03 $\cmsPm$ 0.02 & 1.7 $\cmsPm$ 1.0 & 5 \\
\cline{3-10} &   & \multirow{3}{*}{[600, 750]} & \multirow{3}{*}{0.5} & \x[500, 1000]&\y A07 & 0.04 $\cmsPm$ 0.01 & 0.08 $\cmsPm$ 0.04 & 4.0 $\cmsPm$ 1.6 & 4 \\
&  & & & [1000, 1500] &\y A08 & 0.08 $\cmsPm$ 0.01 & 0.35 $\cmsPm$ 0.08 & 2.8 $\cmsPm$ 1.3 & 5 \\
&  & & &$\geq$1500 &\y A09 & 0.17 $\cmsPm$ 0.02 & 0.02 $\cmsPm$ 0.02 & 1.8 $\cmsPm$ 1.2 & 2 \\
\cline{3-10} &   & $\geq$750\x & 0.5 & $\geq$500\x&\y A10 & 1.01 $\cmsPm$ 0.04 & 0.28 $\cmsPm$ 0.07 & 2.6 $\cmsPm$ 1.1 & 2 \\

\cline{2-10} & \multirow{9}{*}{${=}2$} & \multirow{3}{*}{[250, 450]} & \multirow{3}{*}{1.0} & \x[500, 1000]&\y B01 & 0.01 $\cmsPm$ 0.01 & 2.06 $\cmsPm$ 0.20 & 147 $\cmsPm$ 11\x & 143 \\
&  & & & [1000, 1500] &\y B02 & 0.04 $\cmsPm$ 0.01 & ${<}0.01$ & 43.5 $\cmsPm$ 7.5\x & 37 \\
&  & & &$\geq$1500 &\y B03 & 0.13 $\cmsPm$ 0.01 & ${<}0.01$ & 10.9 $\cmsPm$ 2.8\x & 12 \\
\cline{3-10} &   & \multirow{3}{*}{[450, 600]} & \multirow{3}{*}{0.75} & \x[500, 1000]&\y B04 & 0.02 $\cmsPm$ 0.01 & 0.54 $\cmsPm$ 0.10 & 9.4 $\cmsPm$ 2.2 & 10 \\
&  & & & [1000, 1500] &\y B05 & 0.10 $\cmsPm$ 0.01 & 0.17 $\cmsPm$ 0.06 & 3.4 $\cmsPm$ 1.7 & 9 \\
&  & & & $\geq$1500 &\y B06 & 0.19 $\cmsPm$ 0.02 & ${<}0.01$ & 1.39 $\cmsPm$ 0.82 & 2 \\
\cline{3-10} &   & \multirow{3}{*}{[600, 750]}  & \multirow{3}{*}{0.5} & \x[500, 1000]&\y B07 & 0.03 $\cmsPm$ 0.01 & ${<}0.01$ & 2.4 $\cmsPm$ 1.3 & 3 \\
&  & & & [1000, 1500] &\y B08 & 0.10 $\cmsPm$ 0.01 & 0.26 $\cmsPm$ 0.07 & 1.16 $\cmsPm$ 0.90 & 1 \\
&  & & & $\geq$1500 &\y B09 & 0.24 $\cmsPm$ 0.02 & 0.03 $\cmsPm$ 0.02 & 1.05 $\cmsPm$ 0.78 & 0 \\
\cline{3-10} &   & $\geq$750\x & 0.5 & $\geq$500\x&\y B10 & 1.50 $\cmsPm$ 0.05 & 0.32 $\cmsPm$ 0.08 & 0.42 $\cmsPm$ 0.34 & 3 \\

\cline{2-10} & \multirow{6}{*}{$\geq$3} & \multirow{3}{*}{[250, 450]} & \multirow{3}{*}{1.0} & \x[500, 1000]&\y C01 & 0.01 $\cmsPm$ 0.01 & 1.03 $\cmsPm$ 0.14 & 32.9 $\cmsPm$ 3.3\x & 34 \\
&  & & & [1000, 1500] &\y C02 & 0.06 $\cmsPm$ 0.01 & ${<}0.01$ & 10.6 $\cmsPm$ 2.1\x & 5 \\
&  & & & $\geq$1500 &\y C03 & 0.13 $\cmsPm$ 0.01 & ${<}0.01$ & 2.93 $\cmsPm$ 0.91 & 3 \\
\cline{3-10} &   & \multirow{3}{*}{[450, 600]} & \multirow{3}{*}{0.75} & \x[500, 1000]&\y C04 & 0.03 $\cmsPm$ 0.01 & 0.29 $\cmsPm$ 0.07 & 1.38 $\cmsPm$ 0.50 & 2 \\
&  & & & [1000, 1500] &\y C05 & 0.09 $\cmsPm$ 0.01 & 0.20 $\cmsPm$ 0.06 & 0.72 $\cmsPm$ 0.39 & 1 \\
&  & & & $\geq$1500 &\y C06 & 0.20 $\cmsPm$ 0.02 & ${<}0.01$ & 0.66 $\cmsPm$ 0.45 & 0 \\
\cline{3-10} &   & $\geq$600\x & 0.5 & $\geq$500\x&\y C07 & 1.85 $\cmsPm$ 0.05 & 0.23 $\cmsPm$ 0.06 & 1.66 $\cmsPm$ 0.69 & 2 \\
\hline
\multirow{10}{*}{$\geq$9} & \multirow{4}{*}{${=}1$} & \multirow{2}{*}{[250, 450]} & \multirow{2}{*}{1.0} &  \x[500, 1500]&\y D01 & 0.01 $\cmsPm$ 0.01 & 0.90 $\cmsPm$ 0.12 & 7.9 $\cmsPm$ 1.1 & 7 \\
&  & && $\geq$1500 &\y D02 & 0.03 $\cmsPm$ 0.01 & 0.02 $\cmsPm$ 0.02 & 2.15 $\cmsPm$ 0.67 & 1 \\
\cline{3-10} &   & \multirow{2}{*}{$\geq$450\x} & \multirow{2}{*}{0.75} & \x[500, 1500]&\y D03 & 0.13 $\cmsPm$ 0.01 & 0.72 $\cmsPm$ 0.11 & 1.08 $\cmsPm$ 0.39 & 0 \\
&  & & & $\geq$1500 &\y D04 & 0.38 $\cmsPm$ 0.02 & 0.10 $\cmsPm$ 0.04 & 0.50 $\cmsPm$ 0.27 & 1 \\

\cline{2-10} & \multirow{4}{*}{${=}2$} & \multirow{2}{*}{[250, 450]} & \multirow{2}{*}{1.0} & \x[500, 1500]&\y E01 & 0.02 $\cmsPm$ 0.01 & 1.15 $\cmsPm$ 0.14 & 7.26 $\cmsPm$ 0.97 & 9 \\
&  & & & $\geq$1500 &\y E02 & 0.08 $\cmsPm$ 0.01 & ${<}0.01$ & 2.81 $\cmsPm$ 0.89 & 4 \\
\cline{3-10} &   & \multirow{2}{*}{$\geq$450\x} & \multirow{2}{*}{0.75} & \x[500, 1500]&\y E03 & 0.23 $\cmsPm$ 0.02 & 0.83 $\cmsPm$ 0.12 & 0.71 $\cmsPm$ 0.26 & 2 \\
&  & & & $\geq$1500 &\y E04 & 0.72 $\cmsPm$ 0.03 & 0.20 $\cmsPm$ 0.05 & 0.59 $\cmsPm$ 0.31 & 1 \\

\cline{2-10} & \multirow{4}{*}{$\geq$3} & \multirow{2}{*}{[250, 450]} & \multirow{2}{*}{1.0} & \x[500, 1500]&\y F01 & 0.03 $\cmsPm$ 0.01 & 0.79 $\cmsPm$ 0.11 & 3.55 $\cmsPm$ 0.72 & 3 \\
&  & & & $\geq$1500 &\y F02 & 0.13 $\cmsPm$ 0.01 & ${<}0.01$ & 0.83 $\cmsPm$ 0.35 & 0 \\
\cline{3-10} &   & \multirow{2}{*}{$\geq$450\x} & \multirow{2}{*}{0.75} &\x[500, 1500]&\y F03 & 0.31 $\cmsPm$ 0.02 & 0.26 $\cmsPm$ 0.06 & 0.33 $\cmsPm$ 0.17 & 0 \\
&  & & & $\geq$1500 &\y F04 & 1.04 $\cmsPm$ 0.04 & 0.17 $\cmsPm$ 0.05 & 0.05 $\cmsPm$ 0.05 & 0 \\

\end{tabular}}
\end{table*}

\begin{table*}[!ht]
\centering
\topcaption{Definition of search bins and naming convention in the 0-\PQb
  search. Also given are the \DF values that are used to define the CRs and
  the SRs, the numbers of expected background events with combined statistical
  and systematic uncertainties, the observed numbers of events, and the
expected numbers of signal events in the 0-\PQb search bins.}
\label{tab:0b_results}
\cmsTableResize{
\begin{tabular}{c | c | c | c | l | Z | Z | Z | c}
    \multirow{2}{*}{\njet}     & \LT & \DF& \HT     & \multicolumn{1}{c|}{ Bin} & \multicolumn{2}{c|}{Signal T5qqqqWW ($m_{\PSg}$, $m_{\PSGcz}$) $[$TeV$]$} & \multicolumn{1}{c|}{Predicted} & Observed \\%\hline
 & [GeV] & [rad]  & [GeV] &  \multicolumn{1}{c|}{name} & \multicolumn{1}{c|}{(1.5, 1.0)} & \multicolumn{1}{c|}{(1.9, 0.1)} & \multicolumn{1}{c|}{background} & data \\ \hline
\multirow{9}{*}{$5$}
&\multirow{2}{*}{$\x[250,350]\x$} & \multirow{2}{*}{1.0}
&$[500,750]\x$
 & \y G01
  & 1.82  $\cmsPm$ 0.29 & ${<}0.01$& 102  $\cmsPm$ 48
 & 111 \\
& &
&${\geq}750\x$
 & \y G02
 & 0.21  $\cmsPm$ 0.09 & 0.01  $\cmsPm$ 0.01 & 77  $\cmsPm$ 16
 & 100 \\
\cline{2-9}
&\multirow{2}{*}{$\x[350,450]\x$} & \multirow{2}{*}{1.0}
&$[500,750]\x$
 & \y G03
& 2.25  $\cmsPm$ 0.32 & ${<}0.01$& 24  $\cmsPm$ 15
 & 25 \\
& &
&${\geq}750\x$
 & \y G04
 & 0.29  $\cmsPm$ 0.11 & 0.04  $\cmsPm$ 0.01 & 22.8  $\cmsPm$ 8.3
 & 22 \\
\cline{2-9}
&\multirow{3}{*}{$[450,650]$} & \multirow{3}{*}{0.75}
&$[500,750]\x$
 & \y G05
& 3.02  $\cmsPm$ 0.37 & ${<}0.01$& 14.5  $\cmsPm$ 6.5
 & 17 \\
& &
&$\x[750,1250]\x$
 & \y G06
 & 1.40  $\cmsPm$ 0.25 & 0.04  $\cmsPm$ 0.02 & 12.1  $\cmsPm$ 4.7
 & 10 \\
&&
&${\geq}1250$
 & \y G07
 & 0.08  $\cmsPm$ 0.06 & 0.25  $\cmsPm$ 0.04 & 4.2  $\cmsPm$ 1.7
 & 2 \\
\cline{2-9}
&\multirow{3}{*}{${\geq}650$} & \multirow{3}{*}{0.5}
&$[500,750]\x$
 & \y G08
 & 0.74  $\cmsPm$ 0.18 & 0.01  $\cmsPm$ 0.01 & 2.3  $\cmsPm$ 1.5
 & 5 \\
& &
&$\x[750,1250]\x$
 & \y G09
 & 0.49  $\cmsPm$ 0.15 & 0.12  $\cmsPm$ 0.03 & 5.8  $\cmsPm$ 2.0
 & 6 \\
& &
&${\geq}1250$
 & \y G10
 & 0.14  $\cmsPm$ 0.07 & 1.15  $\cmsPm$ 0.08 & 2.7  $\cmsPm$ 1.3
 & 0 \\
\hline
\multirow{9}{*}{$[6,7]$}
&\multirow{2}{*}{$\x[250,350]\x$} & \multirow{2}{*}{1.0}
&$\x[500,1000]\x$
 & \y H01
  & 3.02  $\cmsPm$ 0.36 & ${<}0.01$& 89  $\cmsPm$ 38
 & 85 \\
& &
&${\geq}1000$
 & \y H02
 & 0.31  $\cmsPm$ 0.10 & 0.09  $\cmsPm$ 0.02 & 30.9  $\cmsPm$ 5.1
 & 33 \\
\cline{2-9}
&\multirow{2}{*}{$\x[350,450]\x$} & \multirow{2}{*}{1.0}
&$\x[500,1000]\x$
 & \y H03
 & 4.13  $\cmsPm$ 0.41 & 0.01  $\cmsPm$ 0.01 & 19  $\cmsPm$ 11
 & 31 \\
& &
&${\geq}1000$
 & \y H04
 & 0.52  $\cmsPm$ 0.14 & 0.14  $\cmsPm$ 0.03 & 9.5  $\cmsPm$ 2.3
 & 8 \\
\cline{2-9}
&\multirow{3}{*}{$\x[450,650]\x$} & \multirow{3}{*}{0.75}
&$[500,750]\x$
 & \y H05
& 3.63  $\cmsPm$ 0.39 & ${<}0.01$& 5.7  $\cmsPm$ 3.3
 & 13 \\
& &
&$\x[750,1250]\x$
 & \y H06
 & 3.79  $\cmsPm$ 0.39 & 0.03  $\cmsPm$ 0.01 & 8.2  $\cmsPm$ 3.2
 & 8 \\
&&
&${\geq}1250$
 &  \y H07
 & 0.36  $\cmsPm$ 0.12 & 0.47  $\cmsPm$ 0.05 & 3.6  $\cmsPm$ 1.8
 & 4 \\
\cline{2-9}
&\multirow{3}{*}{${\geq}650$} & \multirow{3}{*}{0.5}
&$[500,750]\x$
 & \y H08
& 0.89  $\cmsPm$ 0.19 & ${<}0.01$& 0.79  $\cmsPm$ 0.53
 & 3 \\
& &
&$\x[750,1250]\x$
 & \y H09
 & 1.77  $\cmsPm$ 0.26 & 0.15  $\cmsPm$ 0.03 & 3.6  $\cmsPm$ 1.4
 & 5 \\
&&
&${\geq}1250$
 & \y H10
 & 0.83  $\cmsPm$ 0.18 & 2.83  $\cmsPm$ 0.12 & 1.83  $\cmsPm$ 0.86
 & 1 \\
\hline
\multirow{7}{*}{${\geq}8$}
&\multirow{2}{*}{$\x[250,350]\x$} & \multirow{2}{*}{1.0}
&$\x[500,1000]\x$
 & \y I01
  & 0.88  $\cmsPm$ 0.18 & ${<}0.01$& 7.0  $\cmsPm$ 2.8
 & 16 \\
& &
&${\geq}1000$
 & \y I02
 & 0.26  $\cmsPm$ 0.09 & 0.03  $\cmsPm$ 0.01 & 6.3  $\cmsPm$ 1.2
 & 4 \\
\cline{2-9}
&\multirow{2}{*}{$\x[350,450]\x$} & \multirow{2}{*}{1.0}
&$\x[500,1000]\x$
 & \y I03
& 0.55  $\cmsPm$ 0.14 & ${<}0.01$& 1.67  $\cmsPm$ 0.77
 & 3 \\
& &
&${\geq}1000$
 & \y I04
 & 0.72  $\cmsPm$ 0.15 & 0.11  $\cmsPm$ 0.02 & 2.65  $\cmsPm$ 0.89
 & 4 \\
\cline{2-9}
&\multirow{2}{*}{$\x[450,650]\x$} & \multirow{2}{*}{0.75}
&$\x[500,1250]\x$
 & \y I05
 & 2.07  $\cmsPm$ 0.26 & 0.01  $\cmsPm$ 0.01 & 0.63  $\cmsPm$ 0.32
 & 0 \\
&&
&${\geq}1250$
 & \y I06
 & 0.45  $\cmsPm$ 0.12 & 0.3  $\cmsPm$ 0.04 & 0.68  $\cmsPm$ 0.35
 & 1 \\
\cline{2-9}
&\multirow{2}{*}{${\geq}650$} & \multirow{2}{*}{0.5}
&$\x[500,1250]\x$
 & \y I07
 & 0.97  $\cmsPm$ 0.18 & 0.04  $\cmsPm$ 0.01 & 0.27  $\cmsPm$ 0.23
 & 1 \\
&&
&${\geq}1250$
 & \y I08
 & 1.12  $\cmsPm$ 0.18 & 1.37  $\cmsPm$ 0.08 & 0.38  $\cmsPm$ 0.24
 & 1 \\
\end{tabular}}
\end{table*}

\begin{table*}[!h]
\centering
\topcaption{Numbers of expected background events with combined statistical
and systematic uncertainty and the observed numbers
of events in aggregated search bins. The expected number of signal
events for the two corresponding benchmark signals for the multi-\PQb and 0-\PQb
analyses, respectively, are given as well.}
\label{tab:aggregated_regions}
\cmsTableResize
{
\begin{tabular}{c | c | c | c | c | Z | Z | Z | c}

\multirow{2}{*}{\nbtag} & \multirow{2}{*}{\njet}  &  \LT & \DF & \HT                    & \multicolumn{2}{c}{Signal T1tttt ($m_{\PSg}$, $m_{\PSGcz}$) [TeV]}   & \multicolumn{1}{c}{Predicted} & Observed        \\
 &   & [GeV] & [rad] & [GeV]                & \multicolumn{1}{c|}{(1.4, 1.1)}& \multicolumn{1}{c|}{(1.9, 0.1)}           & \multicolumn{1}{c|}{background} & data           \\ \hline
$\geq$1 & $\geq$6 & $\geq$600 & 0.5 & $\geq$1000&  2.66 $\cmsPm$ 0.30 & 7.39 $\cmsPm$ 0.14                                         &11.2  $\cmsPm$  3.6\x  & 13                          \\
$\geq$3 & $\geq$6 & $\geq$600 & 0.5 & $\geq$1000&  0.48 $\cmsPm$ 0.12 & 3.07 $\cmsPm$ 0.09                                         &0.84  $\cmsPm$  0.48 & 1                          \\
$\geq$2 & $\geq$9 & $\geq$450 & 0.75 & $\geq$500\x&  1.35 $\cmsPm$ 0.20 & 2.34 $\cmsPm$ 0.08                                         &1.61  $\cmsPm$  0.43 & 3                          \\
$\geq$2 & $\geq$9 & $\geq$450 & 0.75 & $\geq$1500& 0.37 $\cmsPm$ 0.10 & 1.79 $\cmsPm$ 0.07                                         &0.64  $\cmsPm$  0.33 & 1                          \\
$\geq$3 & $\geq$9 & $\geq$250 & 1.0 & $\geq$500\x&   1.12 $\cmsPm$ 0.19 & 1.33 $\cmsPm$ 0.06                                         &4.58  $\cmsPm$  0.83 & 3                          \\
$\geq$3 & $\geq$9 & $\geq$250 & 1.0 & $\geq$1500 & 0.12 $\cmsPm$ 0.05 & 1.02 $\cmsPm$ 0.05                                         &0.81  $\cmsPm$  0.33 & 0                          \\
$\geq$3 & $\geq$9 & $\geq$450 & 0.75 & $\geq$500\x & 0.41 $\cmsPm$ 0.11 & 1.37 $\cmsPm$ 0.06                                         &0.37  $\cmsPm$  0.17 & 0                          \\
$\geq$3 & $\geq$9 & $\geq$450 & 0.75 & $\geq$1500& 0.17 $\cmsPm$ 0.07 & 1.06 $\cmsPm$ 0.05                                         &0.05  $\cmsPm$  0.05 & 0                          \\ 
\hline 
\hline

  \multicolumn{5}{c|}{}                                      & \multicolumn{2}{c|}{Signal T5qqqqWW ($m_{\PSg}$, $m_{\PSGcz}$) [TeV]}  &          \multicolumn{2}{c}{}             \\
 \multicolumn{5}{c|}{}                                    & \multicolumn{1}{c|}{(1.5, 1.0)}& \multicolumn{1}{c|}{(1.9, 0.1)}            &            \multicolumn{2}{c}{}      \\       
 
\hline

0 & $\geq$5 &  $\geq$650 & 0.5  & $\geq$750\x    &   6.15 $\cmsPm$ 0.57 & 6.29 $\cmsPm$ 0.20   &    18.4 $\cmsPm$ 5.1\x & 14                                                              \\
0 & $\geq$6 &  $\geq$450 & 0.75  & $\geq$500\x   &   16.59 $\cmsPm$ 0.94\x & 5.28 $\cmsPm$ 0.19  &    28.8 $\cmsPm$ 6.8\x & 37                                                                \\ 
0 & $\geq$6 &  $\geq$650 & 0.5  & $\geq$1000   &   4.01 $\cmsPm$ 0.46 & 4.98 $\cmsPm$ 0.18   &    5.1 $\cmsPm$ 1.8& 4                                                                   \\ 
0 & $\geq$7 &  $\geq$450 & 0.75  & $\geq$500\x   &   9.47 $\cmsPm$ 0.71 & 3.54 $\cmsPm$ 0.15   &    9.7 $\cmsPm$ 2.5& 11                                                                  \\ 
0 & $\geq$7 &  $\geq$650 & 0.5  & $\geq$500\x    &   4.28 $\cmsPm$ 0.48 & 3.30 $\cmsPm$ 0.15    &    3.8 $\cmsPm$ 1.2& 4                                                                    \\ 
0 & $\geq$8 &  $\geq$250 & 1.0  & $\geq$1250   &   1.82 $\cmsPm$ 0.31 & 1.71 $\cmsPm$ 0.11   &    7.2 $\cmsPm$ 1.9& 8                                                                    \\ 
\end{tabular}}
\end{table*}

The absence of any significant excess consistent with the SUSY signals
considered in the analysis is used to set limits in the parameter space of the
gluino and lightest neutralino masses. Separate likelihood functions, one for
the multi-\PQb analysis and one for the 0-\PQb analysis, are constructed from
the Poisson probability functions for the CR and SR at both high and low jet
multiplicities.  This includes the $\kappa$ values that correct any residual
differences in the $\Rcs$ transfer factors for regions with different jet
multiplicities. As discussed previously, the values of $\kappa$ are obtained
from simulation, and their uncertainties are incorporated in the likelihood
through log-normal constraints.  The estimated contribution from QCD
events in the CR is also included. A possible signal contamination, which can
be up to $10\%$ for the shown benchmark points,  is taken into account by
including signal terms in the likelihood for both the low-$\njet$ regions
as well as for the low-\DF CR of the search bins. For the 0-\PQb
analysis, the relative contributions from \Wjets and \ttjets events determined
in the fits to the \nbtag distribution in the CR are treated as external
measurements. The correlation between the \Wjets and \ttjets production that is
introduced by such fits is also taken into account.  A ``profile'' likelihood
ratio is used as test statistic. The limits at the 95\% confidence level (CL)
are calculated using the asymptotic formulae~\cite{Cowan:2010js} of the
CL$_\mathrm{s}$ criterion~\cite{Junk1999,ClsCite}.

\begin{figure*}
\centering
\includegraphics[width=\cmsFigWidth]{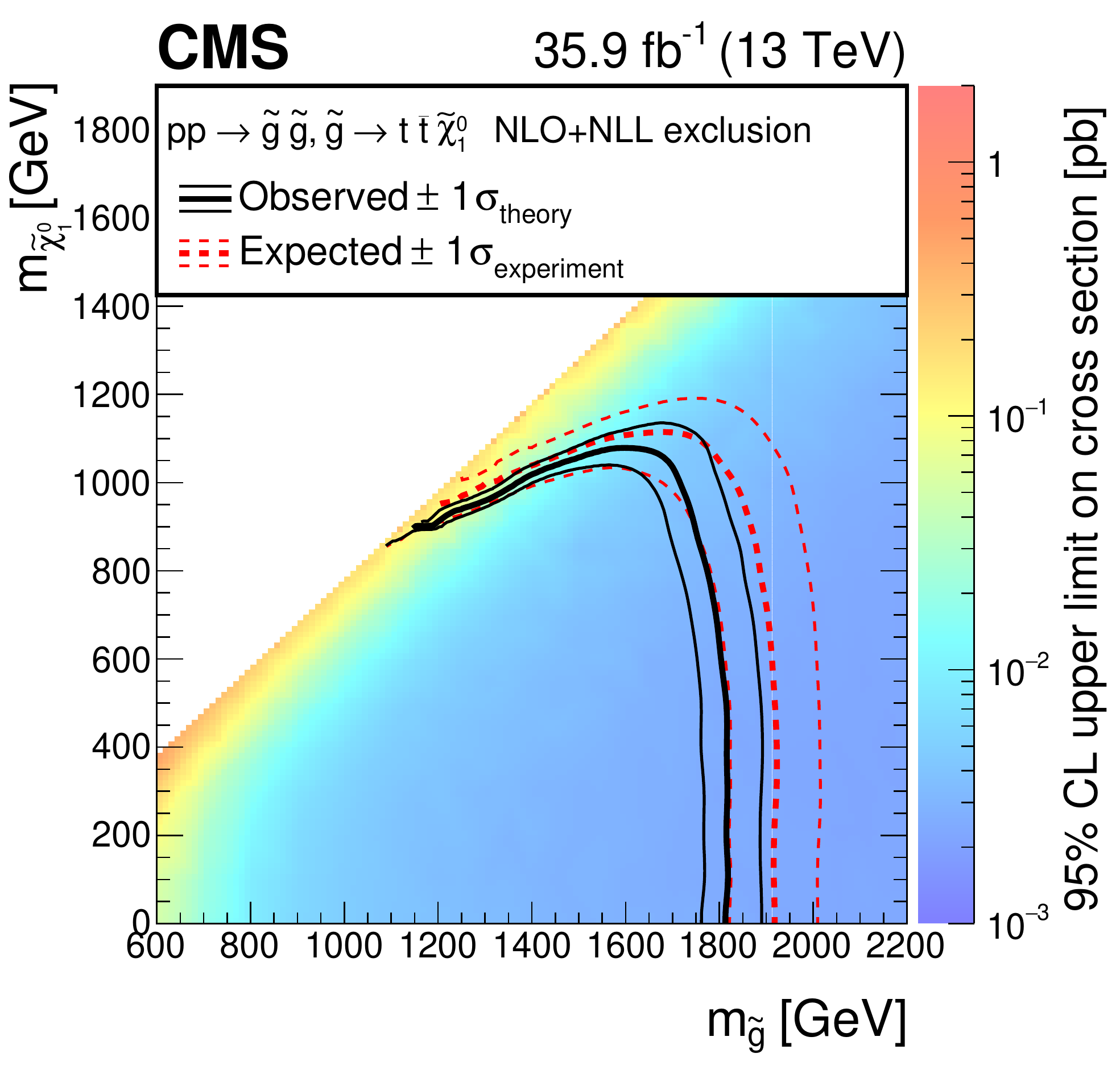} \hfil
\includegraphics[width=\cmsFigWidth]{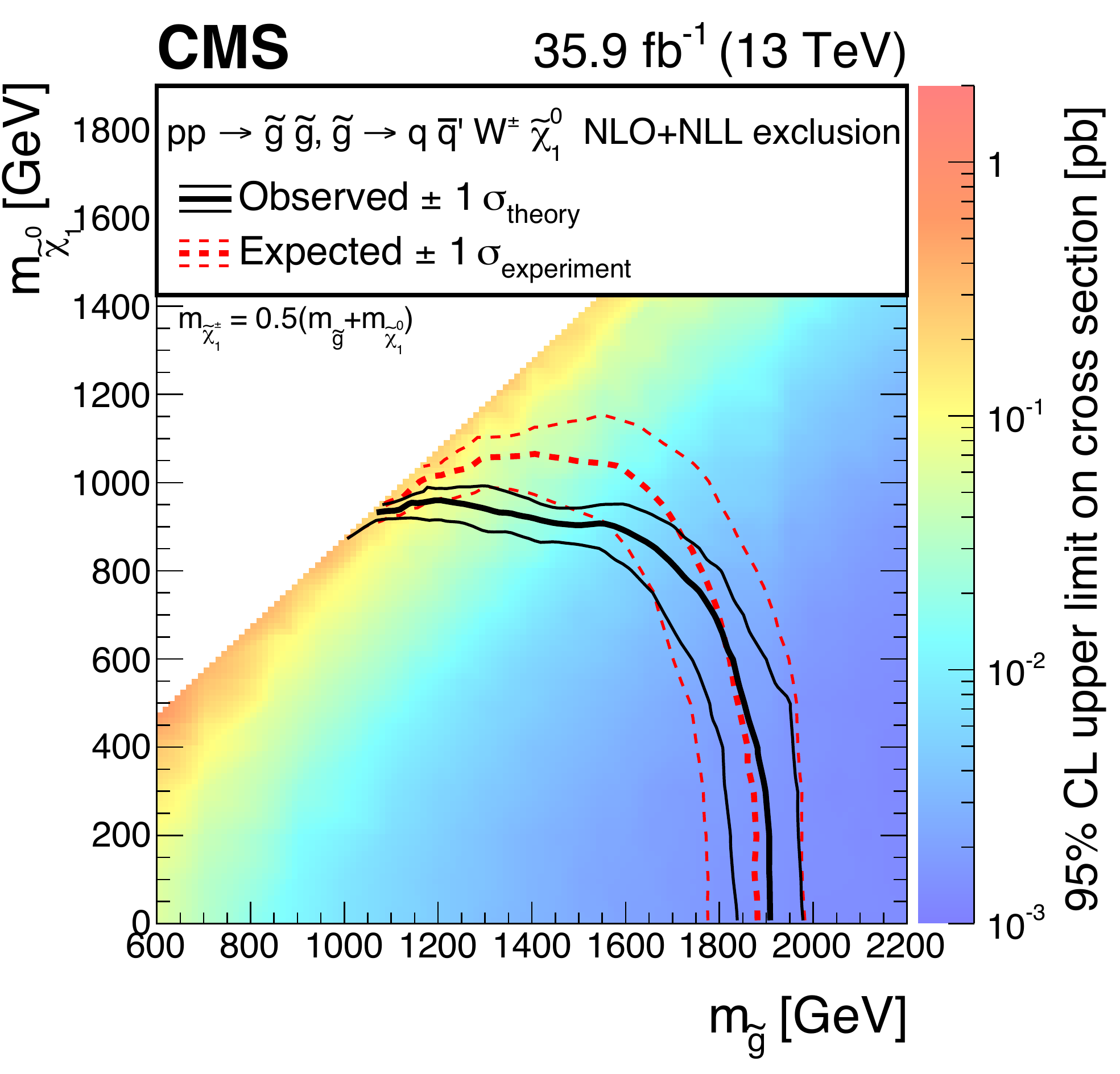}
\caption{Cross section limits at a 95\% CL for the (left) T1tttt and (right)
T5qqqqWW models, as a function of the gluino and LSP masses. In T5qqqqWW, the pair-produced
gluinos decay to first- or second-generation quark-antiquark pairs (\qqbar)
and a chargino (\PSGcpmDo) with its mass taken to be $m_{\PSGcpmDo}=0.5(m_{\PSg}+m_{\PSGczDo})$.
The solid black (dashed red) lines correspond to the observed (expected) mass limits,
with the thicker lines representing the central values and the thinner lines representing
the limits of 68\% uncertainty bands related to the theoretical (experimental) uncertainties.
}
\label{fig:limits}
\end{figure*}

The 95\% CL upper limits on the cross sections, set in the T1tttt model using
the multi-\PQb analysis, and in the T5qqqqWW model using the 0-\PQb analysis,
are shown in Fig.~\ref{fig:limits}. Using the \PSg\PSg\ pair production cross
section calculated at NLO within NLL accuracy, exclusion limits are provided as
a function of the $(m_{\PSg},m_{\PSGczDo})$ mass hypothesis for the data and
for the simulation. For neutralino masses below 800\GeV, gluino masses up to
1.8\TeV are excluded at the 95\% CL in the T1tttt model.  Neutralinos are
excluded up to 1.1\TeV for gluino masses below 1.7\TeV.  In the T5qqqqWW model,
gluino masses up to 1.9\TeV are excluded at the 95\% CL for neutralino masses
below 300\GeV. Neutralinos are excluded up to 950\GeV for gluino masses below
1.2\TeV.

\section{Summary} \label{sec:summary}

A search for supersymmetry has been performed using a 35.9\fbinv sample of
proton-proton collisions at $\sqrt{s}=13\TeV$, recorded by the CMS experiment
in 2016.  Several exclusive search bins are defined that differ in the number
of jets, the number of \PQb-tagged jets, the scalar sum of all jet transverse
momenta as well as the scalar sum of the missing transverse momentum and the
transverse momentum of the lepton.  The main background processes, which arise
from \Wjets and \ttjets in a final state with exactly one lepton and multiple
jets, is reduced significantly by requiring a large azimuthal angle between the
direction of the lepton and of the reconstructed $\PW$ boson, computed under
the hypothesis that all of the missing transverse momentum in the event arises
from a neutrino produced in the leptonic decay of the $\PW$ boson.  The event
yields observed in data are in agreement with the standard model background,
which is estimated using control regions in data and corrections based on
simulation. The lack of any significant excess of events is interpreted in
terms of limits on the parameters of two simplified models that describe gluino
pair production.

For the T1tttt simplified model, in which each gluino decays to a \ttbar pair
and the lightest neutralino, gluino masses up to 1.8\TeV are excluded for
neutralino masses below 800\GeV.  Neutralino masses below 1.1\TeV are excluded
for a gluino mass up to 1.7\TeV. This result extends the exclusion limit from
the previous analysis~\cite{SUS-15-006} on gluino masses by about 250 GeV. The
second simplified model, T5qqqqWW, also describes gluino pair production, but
with decays to first- or second-generation quarks and a chargino, which decays
to a $\PW$ boson and the lightest neutralino. The chargino mass in this decay
channel is assumed to be $m_{\PSGcpmDo}=0.5(m_{\PSg}+m_{\PSGczDo})$. Gluino
masses below 1.9\TeV are excluded for neutralino masses below 300\GeV. This
corresponds to an improvement of about 500 GeV over the previous
result~\cite{SUS-15-006}. For a gluino mass of 1.2\TeV, neutralinos with masses
up to 950\GeV are excluded.

\begin{acknowledgments}

We congratulate our colleagues in the CERN accelerator departments for the excellent performance of the LHC and thank the technical and administrative staffs at CERN and at other CMS institutes for their contributions to the success of the CMS effort. In addition, we gratefully acknowledge the computing centres and personnel of the Worldwide LHC Computing Grid for delivering so effectively the computing infrastructure essential to our analyses. Finally, we acknowledge the enduring support for the construction and operation of the LHC and the CMS detector provided by the following funding agencies: BMWFW and FWF (Austria); FNRS and FWO (Belgium); CNPq, CAPES, FAPERJ, and FAPESP (Brazil); MES (Bulgaria); CERN; CAS, MoST, and NSFC (China); COLCIENCIAS (Colombia); MSES and CSF (Croatia); RPF (Cyprus); SENESCYT (Ecuador); MoER, ERC IUT, and ERDF (Estonia); Academy of Finland, MEC, and HIP (Finland); CEA and CNRS/IN2P3 (France); BMBF, DFG, and HGF (Germany); GSRT (Greece); OTKA and NIH (Hungary); DAE and DST (India); IPM (Iran); SFI (Ireland); INFN (Italy); MSIP and NRF (Republic of Korea); LAS (Lithuania); MOE and UM (Malaysia); BUAP, CINVESTAV, CONACYT, LNS, SEP, and UASLP-FAI (Mexico); MBIE (New Zealand); PAEC (Pakistan); MSHE and NSC (Poland); FCT (Portugal); JINR (Dubna); MON, RosAtom, RAS, RFBR and RAEP (Russia); MESTD (Serbia); SEIDI, CPAN, PCTI and FEDER (Spain); Swiss Funding Agencies (Switzerland); MST (Taipei); ThEPCenter, IPST, STAR, and NSTDA (Thailand); TUBITAK and TAEK (Turkey); NASU and SFFR (Ukraine); STFC (United Kingdom); DOE and NSF (USA).

\hyphenation{Rachada-pisek} Individuals have received support from the Marie-Curie programme and the European Research Council and Horizon 2020 Grant, contract No. 675440 (European Union); the Leventis Foundation; the A. P. Sloan Foundation; the Alexander von Humboldt Foundation; the Belgian Federal Science Policy Office; the Fonds pour la Formation \`a la Recherche dans l'Industrie et dans l'Agriculture (FRIA-Belgium); the Agentschap voor Innovatie door Wetenschap en Technologie (IWT-Belgium); the Ministry of Education, Youth and Sports (MEYS) of the Czech Republic; the Council of Science and Industrial Research, India; the HOMING PLUS programme of the Foundation for Polish Science, cofinanced from European Union, Regional Development Fund, the Mobility Plus programme of the Ministry of Science and Higher Education, the National Science Center (Poland), contracts Harmonia 2014/14/M/ST2/00428, Opus 2014/13/B/ST2/02543, 2014/15/B/ST2/03998, and 2015/19/B/ST2/02861, Sonata-bis 2012/07/E/ST2/01406; the National Priorities Research Program by Qatar National Research Fund; the Programa Severo Ochoa del Principado de Asturias; the Thalis and Aristeia programmes cofinanced by EU-ESF and the Greek NSRF; the Rachadapisek Sompot Fund for Postdoctoral Fellowship, Chulalongkorn University and the Chulalongkorn Academic into Its 2nd Century Project Advancement Project (Thailand); the Welch Foundation, contract C-1845; and the Weston Havens Foundation (USA).

\end{acknowledgments}

\ifthenelse{\boolean{cms@external}}{}{\clearpage}
\bibliography{auto_generated}

\providecommand{\href}[2]{#2}\begingroup\raggedright\begin{thebibliography}{10}%
\makeatletter
\providecommand{\hrefCMSnoop }[0]{\@secondoftwo}%
\makeatother
\providecommand{\doi}{\texttt{doi:}\begingroup \urlstyle{tt}\Url}

\bibitem{Ramond:1971gb}
\hrefCMSnoop {}{P.~Ramond, ``{Dual theory for free fermions}'',} \textit{ Phys.
  Rev. D} \textbf{ 3} (1971) 2415,
\href{http://dx.doi.org/10.1103/PhysRevD.3.2415}{\doi{10.1103/PhysRevD.3.2415}}.
%%CITATION = PHRVA,D3,2415;%%.

\bibitem{Golfand:1971iw}
\hrefCMSnoop {}{{\relax Yu}.~A. Golfand and E.~P. Likhtman, ``{Extension of the
  algebra of {P}oincar\'{e} group generators and violation of {P}
  invariance}'',} \textit{ JETP Lett.} \textbf{ 13} (1971)
323.
%%CITATION = JTPLA,13,323;%%.

\bibitem{Neveu:1971rx}
\hrefCMSnoop {}{A.~Neveu and J.~H. Schwarz, ``{Factorizable dual model of
  pions}'',} \textit{ Nucl. Phys. B} \textbf{ 31} (1971) 86,
\href{http://dx.doi.org/10.1016/0550-3213(71)90448-2}{\doi{10.1016/0550-3213(71)90448-2}}.
%%CITATION = NUPHA,B31,86;%%.

\bibitem{Volkov:1972jx}
\hrefCMSnoop {}{D.~V. Volkov and V.~P. Akulov, ``{Possible universal neutrino
  interaction}'',} \textit{ JETP Lett.} \textbf{ 16} (1972)
438.
%%CITATION = JTPLA,16,438;%%.

\bibitem{Wess:1973kz}
\hrefCMSnoop {}{J.~Wess and B.~Zumino, ``{A {L}agrangian model invariant under
  supergauge transformations}'',} \textit{ Phys. Lett. B} \textbf{ 49} (1974)
  52,
\href{http://dx.doi.org/10.1016/0370-2693(74)90578-4}{\doi{10.1016/0370-2693(74)90578-4}}.
%%CITATION = PHLTA,B49,52;%%.

\bibitem{Wess:1974tw}
\hrefCMSnoop {}{J.~Wess and B.~Zumino, ``{Supergauge transformations in four
  dimensions}'',} \textit{ Nucl. Phys. B} \textbf{ 70} (1974) 39,
\href{http://dx.doi.org/10.1016/0550-3213(74)90355-1}{\doi{10.1016/0550-3213(74)90355-1}}.
%%CITATION = NUPHA,B70,39;%%.

\bibitem{Fayet:1974pd}
\hrefCMSnoop {}{P.~Fayet, ``{Supergauge invariant extension of the {H}iggs
  mechanism and a model for the electron and its neutrino}'',} \textit{ Nucl.
  Phys. B} \textbf{ 90} (1975) 104,
\href{http://dx.doi.org/10.1016/0550-3213(75)90636-7}{\doi{10.1016/0550-3213(75)90636-7}}.
%%CITATION = NUPHA,B90,104;%%.

\bibitem{Nilles:1983ge}
\hrefCMSnoop {}{H.~P. Nilles, ``{Supersymmetry, supergravity and particle
  physics}'',} \textit{ Phys. Rep.} \textbf{ 110} (1984) 1,
\href{http://dx.doi.org/10.1016/0370-1573(84)90008-5}{\doi{10.1016/0370-1573(84)90008-5}}.
%%CITATION = PRPLC,110,1;%%.

\bibitem{Barbieri:1987fn}
\hrefCMSnoop {}{R.~Barbieri and G.~F. Giudice, ``Upper bounds on supersymmetric
  particle masses'',} \textit{ Nucl. Phys. B} \textbf{ 306} (1988) 63,
\href{http://dx.doi.org/10.1016/0550-3213(88)90171-X}{\doi{10.1016/0550-3213(88)90171-X}}.
%%CITATION = NUPHA,B306,63;%%.

\bibitem{FARRAR1978575}
\hrefCMSnoop {}{G.~R. Farrar and P.~Fayet, ``Phenomenology of the production,
  decay, and detection of new hadronic states associated with supersymmetry'',}
  \textit{ Phys. Lett. B} \textbf{ 76} (1978) 575,
  \href{http://dx.doi.org/10.1016/0370-2693(78)90858-4}{\doi{10.1016/0370-2693(78)90858-4}}.

\bibitem{Boehm:1999bj}
\hrefCMSnoop {}{C.~Boehm, A.~Djouadi, and M.~Drees, ``{Light scalar top quarks
  and supersymmetric dark matter}'',} \textit{ Phys. Rev. D} \textbf{ 62}
  (2000) 035012,
  \href{http://dx.doi.org/10.1103/PhysRevD.62.035012}{\doi{10.1103/PhysRevD.62.035012}},
\href{http://www.arXiv.org/abs/hep-ph/9911496}{\texttt{arXiv:hep-ph/9911496}}.
%%CITATION = HEP-PH/9911496;%%.

\bibitem{Balazs:2004bu}
\hrefCMSnoop {}{C.~Balazs, M.~Carena, and C.~E.~M. Wagner, ``{Dark matter,
  light stops and electroweak baryogenesis}'',} \textit{ Phys. Rev. D} \textbf{
  70} (2004) 015007,
  \href{http://dx.doi.org/10.1103/PhysRevD.70.015007}{\doi{10.1103/PhysRevD.70.015007}},
\href{http://www.arXiv.org/abs/hep-ph/0403224}{\texttt{arXiv:hep-ph/0403224}}.
%%CITATION = HEP-PH/0403224;%%.

\bibitem{SUS-15-006}
\hrefCMSnoop {}{{CMS Collaboration}, ``{Search for supersymmetry in events with
  one lepton and multiple jets in proton-proton collisions at $\sqrt{s} =
  13\TeV$}'',} \textit{ Phys. Rev. D} \textbf{ 95} (2017) 012011,
  \href{http://dx.doi.org/10.1103/PhysRevD.95.012011}{\doi{10.1103/PhysRevD.95.012011}},
\href{http://www.arXiv.org/abs/1609.09386}{\texttt{arXiv:1609.09386}}.
%%CITATION = ARXIV:1609.09386;%%.

\bibitem{Chatrchyan:2012ola}
\hrefCMSnoop {}{{CMS Collaboration}, ``{Search for supersymmetry in pp
  collisions at $\sqrt{s}=7\TeV$ in events with a single lepton, jets, and
  missing transverse momentum}'',} \textit{ Eur. Phys. J. C} \textbf{ 73}
  (2013) 2404,
  \href{http://dx.doi.org/10.1140/epjc/s10052-013-2404-z}{\doi{10.1140/epjc/s10052-013-2404-z}},
\href{http://www.arXiv.org/abs/1212.6428}{\texttt{arXiv:1212.6428}}.
%%CITATION = ARXIV:1212.6428;%%.

\bibitem{Chatrchyan:2012sca}
\hrefCMSnoop {}{{CMS Collaboration}, ``{Search for supersymmetry in final
  states with a single lepton, b-quark jets, and missing transverse energy in
  proton-proton collisions at $\sqrt{s}=7\TeV$}'',} \textit{ Phys. Rev. D}
  \textbf{ 87} (2013) 052006,
  \href{http://dx.doi.org/10.1103/PhysRevD.87.052006}{\doi{10.1103/PhysRevD.87.052006}},
\href{http://www.arXiv.org/abs/1211.3143}{\texttt{arXiv:1211.3143}}.
%%CITATION = ARXIV:1211.3143;%%.

\bibitem{Aad:2012ms}
\hrefCMSnoop {}{{ATLAS Collaboration}, ``{Further search for supersymmetry at
  $\sqrt{s}=7\TeV$ in final states with jets, missing transverse momentum and
  isolated leptons with the ATLAS detector}'',} \textit{ Phys. Rev. D} \textbf{
  86} (2012) 092002,
  \href{http://dx.doi.org/10.1103/PhysRevD.86.092002}{\doi{10.1103/PhysRevD.86.092002}},
\href{http://www.arXiv.org/abs/1208.4688}{\texttt{arXiv:1208.4688}}.
%%CITATION = ARXIV:1208.4688;%%.

\bibitem{Chatrchyan:2013iqa}
\hrefCMSnoop {}{{CMS Collaboration}, ``{Search for supersymmetry in pp
  collisions at $\sqrt{s}=8\TeV$ in events with a single lepton, large jet
  multiplicity, and multiple b jets}'',} \textit{ Phys. Lett. B} \textbf{ 733}
  (2014) 328,
  \href{http://dx.doi.org/10.1016/j.physletb.2014.04.023}{\doi{10.1016/j.physletb.2014.04.023}},
\href{http://www.arXiv.org/abs/1311.4937}{\texttt{arXiv:1311.4937}}.
%%CITATION = ARXIV:1311.4937;%%.

\bibitem{Aad:2015mia}
\hrefCMSnoop {}{{ATLAS Collaboration}, ``{Search for squarks and gluinos in
  events with isolated leptons, jets and missing transverse momentum at
  $\sqrt{s}=8\TeV$ with the ATLAS detector}'',} \textit{ JHEP} \textbf{ 04}
  (2015) 116,
  \href{http://dx.doi.org/10.1007/JHEP04(2015)116}{\doi{10.1007/JHEP04(2015)116}},
\href{http://www.arXiv.org/abs/1501.03555}{\texttt{arXiv:1501.03555}}.
%%CITATION = ARXIV:1501.03555;%%.

\bibitem{Aad:2014lra}
\hrefCMSnoop {}{{ATLAS Collaboration}, ``{Search for strong production of
  supersymmetric particles in final states with missing transverse momentum and
  at least three b-jets at $\sqrt{s} = 8\TeV$ proton-proton collisions with the
  ATLAS detector}'',} \textit{ JHEP} \textbf{ 10} (2014) 024,
  \href{http://dx.doi.org/10.1007/JHEP10(2014)024}{\doi{10.1007/JHEP10(2014)024}},
\href{http://www.arXiv.org/abs/1407.0600}{\texttt{arXiv:1407.0600}}.
%%CITATION = ARXIV:1407.0600;%%.

\bibitem{CMS-PAS-SUS-15-007}
\hrefCMSnoop {}{{CMS Collaboration}, ``{Search for supersymmetry in pp
  collisions at $\sqrt{s}=13\TeV$ in the single-lepton final state using the
  sum of masses of large-radius jets}'',} \textit{ JHEP} \textbf{ 08} (2016)
  122,
  \href{http://dx.doi.org/10.1007/JHEP08(2016)122}{\doi{10.1007/JHEP08(2016)122}},
\href{http://www.arXiv.org/abs/1605.04608}{\texttt{arXiv:1605.04608}}.
%%CITATION = ARXIV:1605.04608;%%.

\bibitem{ATLAS-13TeV_single_lepton}
\hrefCMSnoop {}{{ATLAS Collaboration}, ``{Search for gluinos in events with an
  isolated lepton, jets and missing transverse momentum at $\sqrt{s} = 13\TeV$
  with the ATLAS detector}'',} \textit{ Eur. Phys. J. C} \textbf{ 76} (2016)
  565,
  \href{http://dx.doi.org/10.1140/epjc/s10052-016-4397-x}{\doi{10.1140/epjc/s10052-016-4397-x}},
\href{http://www.arXiv.org/abs/1605.04285}{\texttt{arXiv:1605.04285}}.
%%CITATION = ARXIV:1605.04285;%%.

\bibitem{ATLAS-13TeV_multib}
\hrefCMSnoop {}{{ATLAS Collaboration}, ``{Search for pair production of gluinos
  decaying via stop and sbottom in events with $b$-jets and large missing
  transverse momentum in $pp$ collisions at $\sqrt{s} = 13\TeV$ with the ATLAS
  detector}'',} \textit{ Phys. Rev. D} \textbf{ 94} (2016) 032003,
  \href{http://dx.doi.org/10.1103/PhysRevD.94.032003}{\doi{10.1103/PhysRevD.94.032003}},
\href{http://www.arXiv.org/abs/1605.09318}{\texttt{arXiv:1605.09318}}.
%%CITATION = ARXIV:1605.09318;%%.

\bibitem{bib-sms-1}
N.~Arkani-Hamed\hrefCMSnoop {}{ {et~al.}, ``{{MARMOSET}: The path from {LHC}
  data to the new standard model via on-shell effective theories}'',} (2007).
\href{http://www.arXiv.org/abs/hep-ph/0703088}{\texttt{arXiv:hep-ph/0703088}}.
%%CITATION = HEP-PH/0703088;%%.

\bibitem{bib-sms-2}
\hrefCMSnoop {}{J.~Alwall, P.~C. Schuster, and N.~Toro, ``Simplified models for
  a first characterization of new physics at the {LHC}'',} \textit{ Phys. Rev.
  D} \textbf{ 79} (2009) 075020,
  \href{http://dx.doi.org/10.1103/PhysRevD.79.075020}{\doi{10.1103/PhysRevD.79.075020}},
\href{http://www.arXiv.org/abs/0810.3921}{\texttt{arXiv:0810.3921}}.
%%CITATION = ARXIV:0810.3921;%%.

\bibitem{bib-sms-3}
\hrefCMSnoop {}{J.~Alwall, M.-P. Le, M.~Lisanti, and J.~G. Wacker,
  ``{Model-independent jets plus missing energy searches}'',} \textit{ Phys.
  Rev. D} \textbf{ 79} (2009) 015005,
  \href{http://dx.doi.org/10.1103/PhysRevD.79.015005}{\doi{10.1103/PhysRevD.79.015005}},
\href{http://www.arXiv.org/abs/0809.3264}{\texttt{arXiv:0809.3264}}.
%%CITATION = ARXIV:0809.3264;%%.

\bibitem{bib-sms-4}
D.~Alves\hrefCMSnoop {}{ {et~al.}, ``Simplified models for {LHC} new physics
  searches'',} \textit{ J. Phys. G} \textbf{ 39} (2012) 105005,
  \href{http://dx.doi.org/10.1088/0954-3899/39/10/105005}{\doi{10.1088/0954-3899/39/10/105005}},
\href{http://www.arXiv.org/abs/1105.2838}{\texttt{arXiv:1105.2838}}.
%%CITATION = ARXIV:1105.2838;%%.

\bibitem{Chatrchyan:2008zzk}
\hrefCMSnoop {}{{CMS Collaboration}, ``The {CMS} experiment at the {CERN}
  {LHC}'',} \textit{ JINST} \textbf{ 3} (2008) S08004,
\href{http://dx.doi.org/10.1088/1748-0221/3/08/S08004}{\doi{10.1088/1748-0221/3/08/S08004}}.
%%CITATION = JINST,3,S08004;%%.

\bibitem{Sirunyan:2017ulk}
\hrefCMSnoop {}{{CMS Collaboration}, ``{Particle-flow reconstruction and global
  event description with the CMS detector}'',} \textit{ JINST} \textbf{ 12}
  (2017), no.~10, P10003,
  \href{http://dx.doi.org/10.1088/1748-0221/12/10/P10003}{\doi{10.1088/1748-0221/12/10/P10003}},
\href{http://www.arXiv.org/abs/1706.04965}{\texttt{arXiv:1706.04965}}.
%%CITATION = ARXIV:1706.04965;%%.

\bibitem{Cacciari:2007fd}
\hrefCMSnoop {}{M.~Cacciari and G.~P. Salam, ``{Pileup subtraction using jet
  areas}'',} \textit{ Phys. Lett. B} \textbf{ 659} (2008) 119,
  \href{http://dx.doi.org/10.1016/j.physletb.2007.09.077}{\doi{10.1016/j.physletb.2007.09.077}},
\href{http://www.arXiv.org/abs/0707.1378}{\texttt{arXiv:0707.1378}}.
%%CITATION = ARXIV:0707.1378;%%.

\bibitem{Cacciari:2008gp}
\hrefCMSnoop {}{M.~Cacciari, G.~P. Salam, and G.~Soyez, ``{The anti-$k_t$ jet
  clustering algorithm}'',} \textit{ JHEP} \textbf{ 04} (2008) 063,
  \href{http://dx.doi.org/10.1088/1126-6708/2008/04/063}{\doi{10.1088/1126-6708/2008/04/063}},
\href{http://www.arXiv.org/abs/0802.1189}{\texttt{arXiv:0802.1189}}.
%%CITATION = ARXIV:0802.1189;%%.

\bibitem{Chatrchyan:2011ds}
\hrefCMSnoop {}{{CMS Collaboration}, ``{Determination of jet energy calibration
  and transverse momentum resolution in {CMS}}'',} \textit{ JINST} \textbf{ 6}
  (2011) P11002,
  \href{http://dx.doi.org/10.1088/1748-0221/6/11/P11002}{\doi{10.1088/1748-0221/6/11/P11002}},
\href{http://www.arXiv.org/abs/1107.4277}{\texttt{arXiv:1107.4277}}.
%%CITATION = ARXIV:1107.4277;%%.

\bibitem{Cacciari:2011ma}
\hrefCMSnoop {}{M.~Cacciari, G.~P. Salam, and G.~Soyez, ``Fastjet user
  manual'',} \textit{ Eur. Phys. J. C} \textbf{ 72} (2012) 1896,
  \href{http://dx.doi.org/10.1140/epjc/s10052-012-1896-2}{\doi{10.1140/epjc/s10052-012-1896-2}},
\href{http://www.arXiv.org/abs/1111.6097}{\texttt{arXiv:1111.6097}}.
%%CITATION = ARXIV:1111.6097;%%.

\bibitem{Khachatryan:2016kdb}
\hrefCMSnoop {}{{CMS Collaboration}, ``{Jet energy scale and resolution in the
  CMS experiment in pp collisions at 8\TeV}'',} \textit{ JINST} \textbf{ 12}
  (2017) P02014,
  \href{http://dx.doi.org/10.1088/1748-0221/12/02/P02014}{\doi{10.1088/1748-0221/12/02/P02014}},
\href{http://www.arXiv.org/abs/1607.03663}{\texttt{arXiv:1607.03663}}.
%%CITATION = ARXIV:1607.03663;%%.

\bibitem{Chatrchyan:2012jua}
\hrefCMSnoop {}{{CMS Collaboration}, ``{Identification of b-quark jets with the
  CMS experiment}'',} \textit{ JINST} \textbf{ 8} (2013) P04013,
  \href{http://dx.doi.org/10.1088/1748-0221/8/04/P04013}{\doi{10.1088/1748-0221/8/04/P04013}},
\href{http://www.arXiv.org/abs/1211.4462}{\texttt{arXiv:1211.4462}}.
%%CITATION = ARXIV:1211.4462;%%.

\bibitem{Sirunyan:2017ezt}
\hrefCMSnoop {}{{CMS Collaboration}, ``{Identification of heavy-flavour jets
  with the CMS detector in pp collisions at 13 TeV}'',}
\href{http://www.arXiv.org/abs/1712.07158}{\texttt{arXiv:1712.07158}}.
%%CITATION = ARXIV:1712.07158;%%.

\bibitem{Alwall:2014hca}
J.~Alwall\hrefCMSnoop {}{ {et~al.}, ``{The automated computation of tree-level
  and next-to-leading order differential cross sections, and their matching to
  parton shower simulations}'',} \textit{ JHEP} \textbf{ 07} (2014) 079,
  \href{http://dx.doi.org/10.1007/JHEP07(2014)079}{\doi{10.1007/JHEP07(2014)079}},
\href{http://www.arXiv.org/abs/1405.0301}{\texttt{arXiv:1405.0301}}.
%%CITATION = ARXIV:1405.0301;%%.

\bibitem{Nason:2004rx}
\hrefCMSnoop {}{P.~Nason, ``{A new method for combining NLO QCD with shower
  Monte Carlo algorithms}'',} \textit{ JHEP} \textbf{ 11} (2004) 040,
  \href{http://dx.doi.org/10.1088/1126-6708/2004/11/040}{\doi{10.1088/1126-6708/2004/11/040}},
\href{http://www.arXiv.org/abs/hep-ph/0409146}{\texttt{arXiv:hep-ph/0409146}}.
%%CITATION = HEP-PH/0409146;%%.

\bibitem{Frixione:2007vw}
\hrefCMSnoop {}{S.~Frixione, P.~Nason, and C.~Oleari, ``{Matching NLO QCD
  computations with parton shower simulations: the POWHEG method}'',} \textit{
  JHEP} \textbf{ 11} (2007) 070,
  \href{http://dx.doi.org/10.1088/1126-6708/2007/11/070}{\doi{10.1088/1126-6708/2007/11/070}},
\href{http://www.arXiv.org/abs/0709.2092}{\texttt{arXiv:0709.2092}}.
%%CITATION = ARXIV:0709.2092;%%.

\bibitem{Alioli:2010xd}
\hrefCMSnoop {}{S.~Alioli, P.~Nason, C.~Oleari, and E.~Re, ``{A general
  framework for implementing NLO calculations in shower Monte Carlo programs:
  the POWHEG BOX}'',} \textit{ JHEP} \textbf{ 06} (2010) 043,
  \href{http://dx.doi.org/10.1007/JHEP06(2010)043}{\doi{10.1007/JHEP06(2010)043}},
\href{http://www.arXiv.org/abs/1002.2581}{\texttt{arXiv:1002.2581}}.
%%CITATION = ARXIV:1002.2581;%%.

\bibitem{Alioli:2009je}
\hrefCMSnoop {}{S.~Alioli, P.~Nason, C.~Oleari, and E.~Re, ``{NLO single-top
  production matched with shower in POWHEG: $s$- and $t$-channel
  contributions}'',} \textit{ JHEP} \textbf{ 09} (2009) 111,
  \href{http://dx.doi.org/10.1088/1126-6708/2009/09/111}{\doi{10.1088/1126-6708/2009/09/111}},
  \href{http://www.arXiv.org/abs/0907.4076}{\texttt{arXiv:0907.4076}}.
[Erratum: \DOI{10.1007/JHEP02(2010)011}].
%%CITATION = ARXIV:0907.4076;%%.

\bibitem{Re:2010bp}
\hrefCMSnoop {}{E.~Re, ``{Single-top $Wt$-channel production matched with
  parton showers using the POWHEG method}'',} \textit{ Eur. Phys. J. C}
  \textbf{ 71} (2011) 1547,
  \href{http://dx.doi.org/10.1140/epjc/s10052-011-1547-z}{\doi{10.1140/epjc/s10052-011-1547-z}},
\href{http://www.arXiv.org/abs/1009.2450}{\texttt{arXiv:1009.2450}}.
%%CITATION = ARXIV:1009.2450;%%.

\bibitem{Melia:2011tj}
\hrefCMSnoop {}{T.~Melia, P.~Nason, R.~Rontsch, and G.~Zanderighi,
  ``{W$^+$W$^-$, WZ and ZZ production in the POWHEG BOX}'',} \textit{ JHEP}
  \textbf{ 11} (2011) 078,
  \href{http://dx.doi.org/10.1007/JHEP11(2011)078}{\doi{10.1007/JHEP11(2011)078}},
\href{http://www.arXiv.org/abs/1107.5051}{\texttt{arXiv:1107.5051}}.
%%CITATION = ARXIV:1107.5051;%%.

\bibitem{Beneke:2011mq}
\hrefCMSnoop {}{M.~Beneke, P.~Falgari, S.~Klein, and C.~Schwinn, ``{Hadronic
  top-quark pair production with NNLL threshold resummation}'',} \textit{ Nucl.
  Phys. B} \textbf{ 855} (2012) 695,
  \href{http://dx.doi.org/10.1016/j.nuclphysb.2011.10.021}{\doi{10.1016/j.nuclphysb.2011.10.021}},
\href{http://www.arXiv.org/abs/1109.1536}{\texttt{arXiv:1109.1536}}.
%%CITATION = ARXIV:1109.1536;%%.

\bibitem{Cacciari:2011hy}
M.~Cacciari\hrefCMSnoop {}{ {et~al.}, ``{Top-pair production at hadron
  colliders with next-to-next-to-leading logarithmic soft-gluon
  resummation}'',} \textit{ Phys. Lett. B} \textbf{ 710} (2012) 612,
  \href{http://dx.doi.org/10.1016/j.physletb.2012.03.013}{\doi{10.1016/j.physletb.2012.03.013}},
\href{http://www.arXiv.org/abs/1111.5869}{\texttt{arXiv:1111.5869}}.
%%CITATION = ARXIV:1111.5869;%%.

\bibitem{Baernreuther:2012ws}
\hrefCMSnoop {}{P.~B{\"{a}}rnreuther, M.~Czakon, and A.~Mitov, ``Percent level
  precision physics at the tevatron: First genuine {NNLO} {QCD} corrections to
  {$q \bar{q} \to t \bar{t} + X$}'',} \textit{ Phys. Rev. Lett.} \textbf{ 109}
  (2012) 132001,
  \href{http://dx.doi.org/10.1103/PhysRevLett.109.132001}{\doi{10.1103/PhysRevLett.109.132001}},
\href{http://www.arXiv.org/abs/1204.5201}{\texttt{arXiv:1204.5201}}.
%%CITATION = ARXIV:1204.5201;%%.

\bibitem{Czakon:2012zr}
\hrefCMSnoop {}{M.~Czakon and A.~Mitov, ``{NNLO corrections to top-pair
  production at hadron colliders: the all-fermionic scattering channels}'',}
  \textit{ JHEP} \textbf{ 12} (2012) 054,
  \href{http://dx.doi.org/10.1007/JHEP12(2012)054}{\doi{10.1007/JHEP12(2012)054}},
\href{http://www.arXiv.org/abs/1207.0236}{\texttt{arXiv:1207.0236}}.
%%CITATION = ARXIV:1207.0236;%%.

\bibitem{Czakon:2012pz}
\hrefCMSnoop {}{M.~Czakon and A.~Mitov, ``{NNLO corrections to top pair
  production at hadron colliders: the quark-gluon reaction}'',} \textit{ JHEP}
  \textbf{ 01} (2013) 080,
  \href{http://dx.doi.org/10.1007/JHEP01(2013)080}{\doi{10.1007/JHEP01(2013)080}},
\href{http://www.arXiv.org/abs/1210.6832}{\texttt{arXiv:1210.6832}}.
%%CITATION = ARXIV:1210.6832;%%.

\bibitem{Czakon:2013goa}
\hrefCMSnoop {}{M.~Czakon, P.~Fiedler, and A.~Mitov, ``Total top-quark
  pair-production cross section at hadron colliders through
  {$\mathcal{O}(\alpha_S^4)$}'',} \textit{ Phys. Rev. Lett.} \textbf{ 110}
  (2013) 252004,
  \href{http://dx.doi.org/10.1103/PhysRevLett.110.252004}{\doi{10.1103/PhysRevLett.110.252004}},
\href{http://www.arXiv.org/abs/1303.6254}{\texttt{arXiv:1303.6254}}.
%%CITATION = ARXIV:1303.6254;%%.

\bibitem{Gavin:2012sy}
\hrefCMSnoop {}{R.~Gavin, Y.~Li, F.~Petriello, and S.~Quackenbush, ``{W}
  physics at the {LHC} with {FEWZ 2.1}'',} \textit{ Comput. Phys. Commun.}
  \textbf{ 184} (2013) 208,
  \href{http://dx.doi.org/10.1016/j.cpc.2012.09.005}{\doi{10.1016/j.cpc.2012.09.005}},
\href{http://www.arXiv.org/abs/1201.5896}{\texttt{arXiv:1201.5896}}.
%%CITATION = ARXIV:1201.5896;%%.

\bibitem{Gavin:2010az}
\hrefCMSnoop {}{R.~Gavin, Y.~Li, F.~Petriello, and S.~Quackenbush, ``{FEWZ
  2.0}: A code for hadronic {Z} production at next-to-next-to-leading order'',}
  \textit{ Comput. Phys. Commun.} \textbf{ 182} (2011) 2388,
  \href{http://dx.doi.org/10.1016/j.cpc.2011.06.008}{\doi{10.1016/j.cpc.2011.06.008}},
\href{http://www.arXiv.org/abs/1011.3540}{\texttt{arXiv:1011.3540}}.
%%CITATION = ARXIV:1011.3540;%%.

\bibitem{Ball:2014uwa}
\hrefCMSnoop {}{{NNPDF} Collaboration, ``{Parton distributions for the LHC Run
  II}'',} \textit{ JHEP} \textbf{ 04} (2015) 040,
  \href{http://dx.doi.org/10.1007/JHEP04(2015)040}{\doi{10.1007/JHEP04(2015)040}},
\href{http://www.arXiv.org/abs/1410.8849}{\texttt{arXiv:1410.8849}}.
%%CITATION = ARXIV:1410.8849;%%.

\bibitem{Sjostrand:2014zea}
T.~Sj{\"o}strand\hrefCMSnoop {}{ {et~al.}, ``An introduction to {PYTHIA
  8.2}'',} \textit{ Comput. Phys. Commun.} \textbf{ 191} (2015) 159,
  \href{http://dx.doi.org/10.1016/j.cpc.2015.01.024}{\doi{10.1016/j.cpc.2015.01.024}},
\href{http://www.arXiv.org/abs/1410.3012}{\texttt{arXiv:1410.3012}}.
%%CITATION = ARXIV:1410.3012;%%.

\bibitem{bib-nlo-nll-01}
\hrefCMSnoop {}{W.~Beenakker, R.~H{\"o}pker, M.~Spira, and P.~M. Zerwas,
  ``Squark and gluino production at hadron colliders'',} \textit{ Nucl. Phys.
  B} \textbf{ 492} (1997) 51,
  \href{http://dx.doi.org/10.1016/S0550-3213(97)00084-9}{\doi{10.1016/S0550-3213(97)00084-9}},
\href{http://www.arXiv.org/abs/hep-ph/9610490}{\texttt{arXiv:hep-ph/9610490}}.
%%CITATION = HEP-PH/9610490;%%.

\bibitem{bib-nlo-nll-02}
\hrefCMSnoop {}{A.~Kulesza and L.~Motyka, ``Threshold resummation for
  squark-antisquark and gluino-pair production at the {LHC}'',} \textit{ Phys.
  Rev. Lett.} \textbf{ 102} (2009) 111802,
  \href{http://dx.doi.org/10.1103/PhysRevLett.102.111802}{\doi{10.1103/PhysRevLett.102.111802}},
\href{http://www.arXiv.org/abs/0807.2405}{\texttt{arXiv:0807.2405}}.
%%CITATION = ARXIV:0807.2405;%%.

\bibitem{bib-nlo-nll-03}
\hrefCMSnoop {}{A.~Kulesza and L.~Motyka, ``Soft gluon resummation for the
  production of gluino-gluino and squark-antisquark pairs at the {LHC}'',}
  \textit{ Phys. Rev. D} \textbf{ 80} (2009) 095004,
  \href{http://dx.doi.org/10.1103/PhysRevD.80.095004}{\doi{10.1103/PhysRevD.80.095004}},
\href{http://www.arXiv.org/abs/0905.4749}{\texttt{arXiv:0905.4749}}.
%%CITATION = ARXIV:0905.4749;%%.

\bibitem{bib-nlo-nll-04}
W.~Beenakker\hrefCMSnoop {}{ {et~al.}, ``{Soft-gluon resummation for squark and
  gluino hadroproduction}'',} \textit{ JHEP} \textbf{ 12} (2009) 041,
  \href{http://dx.doi.org/10.1088/1126-6708/2009/12/041}{\doi{10.1088/1126-6708/2009/12/041}},
\href{http://www.arXiv.org/abs/0909.4418}{\texttt{arXiv:0909.4418}}.
%%CITATION = ARXIV:0909.4418;%%.

\bibitem{bib-nlo-nll-05}
W.~Beenakker\hrefCMSnoop {}{ {et~al.}, ``{Squark and gluino
  hadroproduction}'',} \textit{ Int. J. Mod. Phys. A} \textbf{ 26} (2011) 2637,
  \href{http://dx.doi.org/10.1142/S0217751X11053560}{\doi{10.1142/S0217751X11053560}},
\href{http://www.arXiv.org/abs/1105.1110}{\texttt{arXiv:1105.1110}}.
%%CITATION = ARXIV:1105.1110;%%.

\bibitem{Khachatryan:2015pea}
\hrefCMSnoop {}{{CMS Collaboration}, ``{Event generator tunes obtained from
  underlying event and multiparton scattering measurements}'',} \textit{ Eur.
  Phys. J. C} \textbf{ 76} (2016) 155,
  \href{http://dx.doi.org/10.1140/epjc/s10052-016-3988-x}{\doi{10.1140/epjc/s10052-016-3988-x}},
\href{http://www.arXiv.org/abs/1512.00815}{\texttt{arXiv:1512.00815}}.
%%CITATION = ARXIV:1512.00815;%%.

\bibitem{Agostinelli:2002hh}
\hrefCMSnoop {}{{GEANT4} Collaboration, ``{GEANT4---a simulation toolkit}'',}
  \textit{ Nucl. Instrum. Meth. A} \textbf{ 506} (2003) 250,
\href{http://dx.doi.org/10.1016/S0168-9002(03)01368-8}{\doi{10.1016/S0168-9002(03)01368-8}}.
%%CITATION = NUIMA,A506,250;%%.

\bibitem{Abdullin:2011zz}
\hrefCMSnoop {}{{CMS Collaboration}, ``{The fast simulation of the CMS detector
  at LHC}'',} \textit{ J. Phys. Conf. Ser.} \textbf{ 331} (2011) 032049,
\href{http://dx.doi.org/10.1088/1742-6596/331/3/032049}{\doi{10.1088/1742-6596/331/3/032049}}.
%%CITATION = 00462,331,032049;%%.

\bibitem{Lester:mt2}
\hrefCMSnoop {}{C.~G. Lester and D.~J. Summers, ``Measuring masses of
  semi-invisibly decaying particle pairs produced at hadron colliders'',}
  \textit{ Phys. Lett. B} \textbf{ 463} (1999) 5,
  \href{http://dx.doi.org/10.1016/S0370-2693(99)00945-4}{\doi{10.1016/S0370-2693(99)00945-4}}.

\bibitem{Khachatryan:2015hwa}
\hrefCMSnoop {}{{CMS Collaboration}, ``{Performance of electron reconstruction
  and selection with the CMS detector in proton-proton collisions at $\sqrt{s}
  = 8\TeV$}'',} \textit{ JINST} \textbf{ 10} (2015) P06005,
  \href{http://dx.doi.org/10.1088/1748-0221/10/06/P06005}{\doi{10.1088/1748-0221/10/06/P06005}},
\href{http://www.arXiv.org/abs/1502.02701}{\texttt{arXiv:1502.02701}}.
%%CITATION = ARXIV:1502.02701;%%.

\bibitem{Chatrchyan:2012xi}
\hrefCMSnoop {}{{CMS Collaboration}, ``{Performance of CMS muon reconstruction
  in pp collision events at $\sqrt{s} = 7\TeV$}'',} \textit{ JINST} \textbf{ 7}
  (2012) P10002,
  \href{http://dx.doi.org/10.1088/1748-0221/7/10/P10002}{\doi{10.1088/1748-0221/7/10/P10002}},
\href{http://www.arXiv.org/abs/1206.4071}{\texttt{arXiv:1206.4071}}.
%%CITATION = ARXIV:1206.4071;%%.

\bibitem{Aaboud:2016mmw}
\hrefCMSnoop {}{{ATLAS Collaboration}, ``Measurement of the inelastic
  proton-proton cross section at {$\sqrt{s}=13\TeV$} with the {ATLAS} detector
  at the {LHC}'',} \textit{ Phys. Rev. Lett.} \textbf{ 117} (2016) 182002,
  \href{http://dx.doi.org/10.1103/PhysRevLett.117.182002}{\doi{10.1103/PhysRevLett.117.182002}},
\href{http://www.arXiv.org/abs/1606.02625}{\texttt{arXiv:1606.02625}}.
%%CITATION = ARXIV:1606.02625;%%.

\bibitem{Khachatryan:2016ipq}
\hrefCMSnoop {}{{CMS Collaboration}, ``{Measurement of the production cross
  section of a W boson in association with two b jets in pp collisions at
  $\sqrt{s} = 8\TeV$}'',} \textit{ Eur. Phys. J. C} \textbf{ 77} (2017) 92,
  \href{http://dx.doi.org/10.1140/epjc/s10052-016-4573-z}{\doi{10.1140/epjc/s10052-016-4573-z}},
\href{http://www.arXiv.org/abs/1608.07561}{\texttt{arXiv:1608.07561}}.
%%CITATION = ARXIV:1608.07561;%%.

\bibitem{Bern:2011ie}
Z.~Bern\hrefCMSnoop {}{ {et~al.}, ``{Left-handed W bosons at the LHC}'',}
  \textit{ Phys. Rev. D} \textbf{ 84} (2011) 034008,
  \href{http://dx.doi.org/10.1103/PhysRevD.84.034008}{\doi{10.1103/PhysRevD.84.034008}},
\href{http://www.arXiv.org/abs/1103.5445}{\texttt{arXiv:1103.5445}}.
%%CITATION = ARXIV:1103.5445;%%.

\bibitem{Khachatryan:2015paa}
\hrefCMSnoop {}{{CMS Collaboration}, ``{Angular coefficients of Z bosons
  produced in pp collisions at $\sqrt{s}= 8\TeV$ and decaying to $\mu^+ \mu^-$
  as a function of transverse momentum and rapidity}'',} \textit{ Phys. Lett.
  B} \textbf{ 750} (2015) 154,
  \href{http://dx.doi.org/10.1016/j.physletb.2015.08.061}{\doi{10.1016/j.physletb.2015.08.061}},
\href{http://www.arXiv.org/abs/1504.03512}{\texttt{arXiv:1504.03512}}.
%%CITATION = ARXIV:1504.03512;%%.

\bibitem{Chatrchyan:2011ig}
\hrefCMSnoop {}{{CMS Collaboration}, ``Measurement of the polarization of {W}
  bosons with large transverse momenta in {W}+jets events at the {LHC}'',}
  \textit{ Phys. Rev. Lett.} \textbf{ 107} (2011) 021802,
  \href{http://dx.doi.org/10.1103/PhysRevLett.107.021802}{\doi{10.1103/PhysRevLett.107.021802}},
\href{http://www.arXiv.org/abs/1104.3829}{\texttt{arXiv:1104.3829}}.
%%CITATION = ARXIV:1104.3829;%%.

\bibitem{ATLAS:2012au}
\hrefCMSnoop {}{{ATLAS Collaboration}, ``{Measurement of the polarisation of
  $W$ bosons produced with large transverse momentum in $pp$ collisions at
  $\sqrt{s}=7\TeV$ with the ATLAS experiment}'',} \textit{ Eur. Phys. J. C}
  \textbf{ 72} (2012) 2001,
  \href{http://dx.doi.org/10.1140/epjc/s10052-012-2001-6}{\doi{10.1140/epjc/s10052-012-2001-6}},
\href{http://www.arXiv.org/abs/1203.2165}{\texttt{arXiv:1203.2165}}.
%%CITATION = ARXIV:1203.2165;%%.

\bibitem{Aad:2012ky}
\hrefCMSnoop {}{{ATLAS Collaboration}, ``{Measurement of the $W$ boson
  polarization in top quark decays with the ATLAS detector}'',} \textit{ JHEP}
  \textbf{ 06} (2012) 088,
  \href{http://dx.doi.org/10.1007/JHEP06(2012)088}{\doi{10.1007/JHEP06(2012)088}},
\href{http://www.arXiv.org/abs/1205.2484}{\texttt{arXiv:1205.2484}}.
%%CITATION = ARXIV:1205.2484;%%.

\bibitem{Khachatryan:2015dzz}
\hrefCMSnoop {}{{CMS Collaboration}, ``{Measurement of top quark polarisation
  in t-channel single top quark production}'',} \textit{ JHEP} \textbf{ 04}
  (2016) 073,
  \href{http://dx.doi.org/10.1007/JHEP04(2016)073}{\doi{10.1007/JHEP04(2016)073}},
\href{http://www.arXiv.org/abs/1511.02138}{\texttt{arXiv:1511.02138}}.
%%CITATION = ARXIV:1511.02138;%%.

\bibitem{Aaboud:2016hsq}
\hrefCMSnoop {}{{ATLAS Collaboration}, ``{Measurement of the $W$ boson
  polarisation in $t\bar{t}$ events from pp collisions at $\sqrt{s} = 8\TeV$ in
  the lepton+jets channel with ATLAS}'',} \textit{ Eur. Phys. J. C} \textbf{
  77} (2017) 264,
  \href{http://dx.doi.org/10.1140/epjc/s10052-017-4819-4}{\doi{10.1140/epjc/s10052-017-4819-4}},
\href{http://www.arXiv.org/abs/1612.02577}{\texttt{arXiv:1612.02577}}.
%%CITATION = ARXIV:1612.02577;%%.

\bibitem{Czarnecki:2010gb}
\hrefCMSnoop {}{A.~Czarnecki, J.~G. Korner, and J.~H. Piclum, ``{Helicity
  fractions of $W$ bosons from top quark decays at NNLO in QCD}'',} \textit{
  Phys. Rev. D} \textbf{ 81} (2010) 111503,
  \href{http://dx.doi.org/10.1103/PhysRevD.81.111503}{\doi{10.1103/PhysRevD.81.111503}},
\href{http://www.arXiv.org/abs/1005.2625}{\texttt{arXiv:1005.2625}}.
%%CITATION = ARXIV:1005.2625;%%.

\bibitem{CMS-PAS-LUM-15-001}
\href {http://cdsweb.cern.ch/record/2138682}{{CMS Collaboration}, ``{CMS}
  luminosity measurement for the 2015 data taking period'',} CMS Physics
  Analysis Summary CMS-PAS-LUM-15-001, CERN, 2016.

\bibitem{CMS-PAS-LUM-17-001}
\href {http://cdsweb.cern.ch/record/2257069}{{CMS Collaboration}, ``{CMS}
  luminosity measurements for the 2016 data taking period'',} CMS Physics
  Analysis Summary CMS-PAS-LUM-17-001, CERN, 2017.

\bibitem{Cowan:2010js}
\hrefCMSnoop {}{G.~Cowan, K.~Cranmer, E.~Gross, and O.~Vitells, ``Asymptotic
  formulae for likelihood-based tests of new physics'',} \textit{ Eur. Phys. J.
  C} \textbf{ 71} (2011) 1554,
  \href{http://dx.doi.org/10.1140/epjc/s10052-011-1554-0}{\doi{10.1140/epjc/s10052-011-1554-0}},
  \href{http://www.arXiv.org/abs/1007.1727}{\texttt{arXiv:1007.1727}}.
[Erratum: \DOI{10.1140/epjc/s10052-013-2501-z}].
%%CITATION = ARXIV:1007.1727;%%.

\bibitem{Junk1999}
\hrefCMSnoop {}{T.~Junk, ``{Confidence level computation for combining searches
  with small statistics}'',} \textit{ Nucl. Instrum. Meth. A} \textbf{ 434}
  (1999) 435,
  \href{http://dx.doi.org/10.1016/S0168-9002(99)00498-2}{\doi{10.1016/S0168-9002(99)00498-2}},
\href{http://www.arXiv.org/abs/hep-ex/9902006}{\texttt{arXiv:hep-ex/9902006}}.
%%CITATION = HEP-EX/9902006;%%.

\bibitem{ClsCite}
\hrefCMSnoop {}{A.~L. Read, ``{Presentation of search results: the {$\rm CL_s$}
  technique}'',} \textit{ J. Phys. G} \textbf{ 28} (2002) 2693,
\href{http://dx.doi.org/10.1088/0954-3899/28/10/313}{\doi{10.1088/0954-3899/28/10/313}}.
%%CITATION = JPHGB,G28,2693;%%.

\end{thebibliography}\endgroup
\cleardoublepage \appendix\section{The CMS Collaboration \label{app:collab}}\begin{sloppypar}\hyphenpenalty=5000\widowpenalty=500\clubpenalty=5000\textbf{Yerevan Physics Institute,  Yerevan,  Armenia}\\*[0pt]
A.M.~Sirunyan, A.~Tumasyan
\vskip\cmsinstskip
\textbf{Institut f\"{u}r Hochenergiephysik,  Wien,  Austria}\\*[0pt]
W.~Adam, F.~Ambrogi, E.~Asilar, T.~Bergauer, J.~Brandstetter, E.~Brondolin, M.~Dragicevic, J.~Er\"{o}, M.~Flechl, M.~Friedl, R.~Fr\"{u}hwirth\cmsAuthorMark{1}, V.M.~Ghete, J.~Grossmann, J.~Hrubec, M.~Jeitler\cmsAuthorMark{1}, A.~K\"{o}nig, N.~Krammer, I.~Kr\"{a}tschmer, D.~Liko, T.~Madlener, I.~Mikulec, E.~Pree, D.~Rabady, N.~Rad, H.~Rohringer, J.~Schieck\cmsAuthorMark{1}, R.~Sch\"{o}fbeck, M.~Spanring, D.~Spitzbart, J.~Strauss, W.~Waltenberger, J.~Wittmann, C.-E.~Wulz\cmsAuthorMark{1}, M.~Zarucki
% ...
\vskip\cmsinstskip
% ...
1:~~Also at Vienna University of Technology, Vienna, Austria\\
% ...

\end{sloppypar}
\end{document}